%% file: GSO.tex
\documentclass[dvips]{acm_proc_article-sp}
\input{macros-GSO.tex}

\begin{document}

\title{Gaming security by obscurity}

\numberofauthors{1}
\author{\alignauthor
Dusko Pavlovic\\ 
\affaddr{Royal Holloway, University of London, and University of Twente}\\
\email{dusko.pavlovic@rhul.ac.uk}}

\date{}


\maketitle

\begin{abstract}
Shannon  \cite{ShannonC:Secrecy} sought security against the attacker with unlimited computational powers: \emph{if an information source conveys some information, then Shannon's attacker will surely extract that information}. Diffie and Hellman \cite{Diffie-Hellman} refined Shannon's attacker model by taking into account the fact that the real attackers are computationally limited.
This idea became one of the greatest new paradigms in computer science, and led to modern cryptography. 

Shannon also sought security against the attacker with unlimited logical and observational powers, expressed through the maxim that "the enemy knows the system". This view is still endorsed in cryptography. The popular formulation, going back to Kerckhoffs
\cite{Kerckhoffs}, is that "there is no security by obscurity", meaning that the algorithms cannot be kept obscured from the attacker, and that security should only rely upon the secret keys. In fact, modern cryptography goes even further than Shannon or Kerckhoffs in tacitly assuming that \emph{if there is an algorithm that can break the system, then the attacker will surely find that algorithm}. The attacker is not viewed as an omnipotent computer any more, but he is still construed as an omnipotent programmer. The ongoing hackers' successes seem to justify this view.

So the Diffie-Hellman step from unlimited to limited computational powers has not been extended into a step from unlimited to limited logical or programming powers.  Is the assumption that all feasible algorithms will eventually be discovered and implemented really different from the assumption that everything that is computable will eventually be computed? The present paper explores some ways to refine the current models of the attacker, and of the defender, by taking into account their limited logical and programming powers. If the adaptive attacker actively queries the system to seek out its vulnerabilities, can the system gain some security by actively learning attacker's methods, and adapting to them? 
\end{abstract}

%
%
%

\section{Introduction}
New paradigms change the world. In computer science, they often sneak behind researchers' backs: the grand visions often frazzle into minor ripples (like the fifth generation of programming languages), whereas some modest goals engender tidal waves with global repercussions (like moving the cursor by a device with wheels, or connecting remote computers). So it is not easy to conjure a new paradigm when you need it. 

Perhaps the only readily available method to generate new paradigms at leisure is by disputing the obvious. Just in case, I question on this occasion not one, but two generally endorsed views:
\begin{itemize}
\item\emph{Kerckhoffs Principle} that there is no security by obscurity, and
\item\emph{Fortification Principle} that the defender has to defend all attack vectors, whereas the attacker only needs to attack one.
\end{itemize}
To simplify things a little, I argue that these two principles are related. Kerckhoffs's Principle demands that a system should withstand attackers' unhindered probing. In the modern security definitions, this is amplified to the requirement that the system should resist a family of attacks, irrespective of the details of their algorithms. The adaptive attackers are thus allowed to query the system, whereas the system is not allowed to query the attackers. The resulting \emph{information asymmetry} makes security look like a game biased in favor of the attackers. The Fortification Principle is an expression of that asymmetry. In economics, information asymmetry has been recognized as a fundamental problem, worth the Nobel Prize in Economics for 2001 \cite{Stigler,Akerlof,Spence}. In security research, the problem does not seem to have been explicitly addressed, but there is, of course, no shortage of efforts to realize security by obscurity in practice --- albeit without any discernible method. Although the practices of analyzing the attackers and hiding the systems are hardly waiting for anyone to invent a new paradigm, I will pursue the possibility that a new paradigm might be sneaking behind our backs again, like so many old paradigms did.

\medskip
\subsubsection*{Outline of the paper} 
While I am on the subject of security paradigms, I decided to first spell out a general overview of the old ones.   An attempt at this is in Sec.~\ref{Old}. It is surely incomplete, and perhaps wrongheaded, but it may help a little. It is difficult to communicate about the new without  an agreement about the old. Moreover, it will be interesting to hear not only whether my  new paradigms are new, but also whether my old paradigms are old.  

The new security paradigm arising from the slogan \emph{"Know your enemy"} is discussed in Sec.~\ref{New}. Of course, security engineers often know their enemies, so this is not much of a new paradigm in practice. But security researchers often require that systems should be secure against universal families of attackers, without knowing anything about who the enemy is at any particular moment. With respect to such static requirements, a game theoretic analysis of dynamics of security can be viewed as an almost-new paradigm (with few previous owners). In Sec.\ref{Fortification} I point to the practical developments that lead up to this paradigm, and then in Sec.~\ref{Game} I describe the game of attack vectors, which illustrates it. This is a very crude view of security process as a game of incomplete information. I provide a simple pictorial analysis of the strategic interactions in this game, which turn out to be based on acquiring information about the opponent's type and behavior. A sketch of a formal model of security games of incomplete information, and of the game of attack vectors, is given in Appendix~\ref{Formalism}. 

A brand new security paradigm of \emph{"Applied security by obscurity"} is described in Sec.~\ref{brand}. It is based on the idea of \emph{logical complexity} of programs, which leads to one-way programming similarly like computational complexity led to one-way computations. If achieved, one-way programming will be a powerful tool in security games.

A final attempt at a summary, and some comments about the future research, and the pitfalls, are given in Sec.~\ref{Summary}.

\medskip
\subsubsection*{Related work} 
The two new paradigms offer two new tools for the security toolkit: games of incomplete information, and algorithmic information theory.

Game theoretic techniques have been used in applied security for a long time, since the need for strategic reasoning often arises in practice. A typical example from the early days is \cite{DeanW:pitfalls}, where games of imperfect information were used.  Perhaps the simplest more recent game based model are the attack-defense trees, which boil down to zero-sum extensive games \cite{MauwS:FAST10}. Another application of games of imperfect information appeared, e.g., in a previous edition of this conference  \cite{MooreT:CyberWarrior}. Conspicuously, games of incomplete information do not seem to have been used, which seems appropriate since they analyze how players keep each other in obscurity. The coalgebraic presentation of games and response relations, presented in the Appendix, is closely related with the formalism used in \cite{PavlovicD:CALCO09}. 

The concept of logical complexity, proposed in Sec.~\ref{brand}, is based on the ideas of algorithmic information theory  \cite{Kolmogorov,SolomonoffR:inference} in general, and in particular on the idea of \emph{logical depth} \cite{Chaitin,LevinL:conservation,BennettC:depth}. I propose to formalize logical complexity by lifting logical depth from the G\"odel-Kleene indices to program specifications \cite{Hoare:ProgPred,Bjorner,BorgerE:book-ASM,FiadeiroJ:book,PavlovicD:FOPS,PavlovicD:AMAST02}. The underlying idea that a G\"odel-Kleene index of a program can be viewed as its "explanation" goes back to Kleene's idea of \emph{realizability} \cite{KleeneSC:realizability} and to Solomonoff's formalization of inductive inference \cite{SolomonoffR:inference}.

\input{1-Engineering.tex}
\input{2-Science.tex}

\input{3-Conclusion.tex}

\bibliography{ref-GSO,PavlovicD,games}
\bibliographystyle{plain}

\bigskip
\appendix
\input{4-Appendix.tex}


\end{document}

%% file: macros-GSO.tex
\usepackage{xy}
\xyoption{all}
 \xyoption{line}
\CompileMatrices

\usepackage{amssymb}
\usepackage{amstext}
\usepackage{amsmath}
\usepackage{stmaryrd}
\usepackage{pgf}
\usepackage{color}
\usepackage{pstricks}
\usepackage{array}

\input{prooftree}

\newcommand{\subparagraph}[1]{\medskip\noindent\underline{\it #1}}
\newcommand{\conv}{{\rm conv}}

\newcommand{\System}{{\bf \footnotesize Defense}}
\newcommand{\Attack}{{\color{white} \bf \footnotesize Attack}}

\newcommand{\tto}[1]{\xrightarrow{#1}}
\newcommand{\ttto}[2]{\xrightarrow[#2]{#1}}

\newdef{defn}{Definition}

\newcommand{\CCC}{{\cal C}}
\newcommand{\DDD}{{\cal D}}

\newcommand{\OOO}{{\cal O}}

\newcommand{\RRR}{{\cal R}}

\renewcommand{\Bbb}{\mathbb}

\newcommand{\MMm}{{\Bbb M}}
\newcommand{\NNn}{{\Bbb N}}

\newcommand{\RRr}{{\Bbb R}}

\newcommand{\ZZz}{{\Bbb Z}}

\newcounter{countroman}
\newenvironment{rnumerate}%
{\begin{list}{{\rm (\roman{countroman})}}{\usecounter{countroman}}}%
{\end{list}}

\newcounter{countalpha}
\newenvironment{anumerate}%
{\begin{list}{(\alph{countalpha})}{\usecounter{countalpha}}}%
{\end{list}}

\newcounter{countalphabf}
{\protect\begin{list}{{\rm (}{\bf \protect\alph{countalphabf}}{\rm%
)}}{\protect\usecounter{countalphabf}}}%
{\end{list}}

\mathcode`\<="4268 
\mathcode`\>="5269 
\mathchardef\gt="313E 
\mathchardef\lt="313C 

\newcommand{\be}[1]{\begin{#1}}

\newcommand{\beq}{\begin{equation}}
\newcommand{\eeq}{\end{equation}}
\newcommand{\ba}[1]{\begin{array}{#1}}
\newcommand{\ea}{\end{array}}
\newcommand{\bea}{\begin{eqnarray}}
\newcommand{\eea}{\end{eqnarray}}
\newcommand{\bear}{\begin{eqnarray*}}
\newcommand{\eear}{\end{eqnarray*}}

\renewcommand{\paragraph}[1]{\smallskip\noindent {\textbf{#1}.}}

\newcommand{\Algs}[2]{{\sf RA}({#1},{#2})}
\newcommand{\Al}{{\sf RA}}

\newcommand{\SR}{{\sf SR}}
\newcommand{\MR}{{\sf MR}}

\newcommand{\enc}[1]{\left\ulcorner{#1}\right\urcorner}
\newcommand{\dec}[1]{\left\{{#1}\right\}}

%% file: prooftree.tex
\newdimen\proofrulebreadth \proofrulebreadth=.05em
\newdimen\proofdotseparation \proofdotseparation=1.25ex
\newdimen\proofrulebaseline \proofrulebaseline=2ex
\newcount\proofdotnumber \proofdotnumber=3
\let\then\relax
\def\hfi{\hskip0pt plus.0001fil}
\mathchardef\squigto="3A3B
\newif\ifinsideprooftree\insideprooftreefalse
\newif\ifonleftofproofrule\onleftofproofrulefalse
\newif\ifproofdots\proofdotsfalse
\newif\ifdoubleproof\doubleprooffalse
\let\wereinproofbit\relax
\newdimen\shortenproofleft
\newdimen\shortenproofright
\newdimen\proofbelowshift
\newbox\proofabove
\newbox\proofbelow
\newbox\proofrulename
\def\shiftproofbelow{\let\next\relax\afterassignment\setshiftproofbelow\dimen0 }
\def\shiftproofbelowneg{\def\next{\multiply\dimen0 by-1 }%
\afterassignment\setshiftproofbelow\dimen0 }
\def\setshiftproofbelow{\next\proofbelowshift=\dimen0 }
\def\setproofrulebreadth{\proofrulebreadth}

\def\prooftree{
\ifnum  \lastpenalty=1
\then   \unpenalty
\else   \onleftofproofrulefalse
\fi
\ifonleftofproofrule
\else   \ifinsideprooftree
        \then   \hskip.5em plus1fil
        \fi
\fi
\bgroup
\setbox\proofbelow=\hbox{}\setbox\proofrulename=\hbox{}%
\let\justifies\proofover\let\leadsto\proofoverdots\let\Justifies\proofoverdbl
\let\using\proofusing\let\[\prooftree
\ifinsideprooftree\let\]\endprooftree\fi
\proofdotsfalse\doubleprooffalse
\let\thickness\setproofrulebreadth
\let\shiftright\shiftproofbelow \let\shift\shiftproofbelow
\let\shiftleft\shiftproofbelowneg
\let\ifwasinsideprooftree\ifinsideprooftree
\insideprooftreetrue
\setbox\proofabove=\hbox\bgroup$\displaystyle 
\let\wereinproofbit\prooftree
\shortenproofleft=0pt \shortenproofright=0pt \proofbelowshift=0pt
\onleftofproofruletrue\penalty1
}

\def\eproofbit{
\ifx    \wereinproofbit\prooftree
\then   \ifcase \lastpenalty
        \then   \shortenproofright=0pt  
        \or     \unpenalty\hfil         
        \or     \unpenalty\unskip       
        \else   \shortenproofright=0pt  
        \fi
\fi
\global\dimen0=\shortenproofleft
\global\dimen1=\shortenproofright
\global\dimen2=\proofrulebreadth
\global\dimen3=\proofbelowshift
\global\dimen4=\proofdotseparation
\global\count255=\proofdotnumber
$\egroup  
\shortenproofleft=\dimen0
\shortenproofright=\dimen1
\proofrulebreadth=\dimen2
\proofbelowshift=\dimen3
\proofdotseparation=\dimen4
\proofdotnumber=\count255
}

\def\proofover{
\eproofbit 
\setbox\proofbelow=\hbox\bgroup 
\let\wereinproofbit\proofover
$\displaystyle
}%
\def\proofoverdbl{
\eproofbit 
\doubleprooftrue
\setbox\proofbelow=\hbox\bgroup 
\let\wereinproofbit\proofoverdbl
$\displaystyle
}%
\def\proofoverdots{
\eproofbit 
\proofdotstrue
\setbox\proofbelow=\hbox\bgroup 
\let\wereinproofbit\proofoverdots
$\displaystyle
}%
\def\proofusing{
\eproofbit 
\setbox\proofrulename=\hbox\bgroup 
\let\wereinproofbit\proofusing
\kern0.3em$
}

\def\endprooftree{
\eproofbit 
  \dimen5 =0pt
\dimen0=\wd\proofabove \advance\dimen0-\shortenproofleft
\advance\dimen0-\shortenproofright
\dimen1=.5\dimen0 \advance\dimen1-.5\wd\proofbelow
\dimen4=\dimen1
\advance\dimen1\proofbelowshift \advance\dimen4-\proofbelowshift
\ifdim  \dimen1<0pt
\then   \advance\shortenproofleft\dimen1
        \advance\dimen0-\dimen1
        \dimen1=0pt
        \ifdim  \shortenproofleft<0pt
        \then   \setbox\proofabove=\hbox{%
                        \kern-\shortenproofleft\unhbox\proofabove}%
                \shortenproofleft=0pt
        \fi
\fi
\ifdim  \dimen4<0pt
\then   \advance\shortenproofright\dimen4
        \advance\dimen0-\dimen4
        \dimen4=0pt
\fi
\ifdim  \shortenproofright<\wd\proofrulename
\then   \shortenproofright=\wd\proofrulename
\fi
\dimen2=\shortenproofleft \advance\dimen2 by\dimen1
\dimen3=\shortenproofright\advance\dimen3 by\dimen4
\ifproofdots
\then
        \dimen6=\shortenproofleft \advance\dimen6 .5\dimen0
        \setbox1=\vbox to\proofdotseparation{\vss\hbox{$\cdot$}\vss}%
        \setbox0=\hbox{%
                \advance\dimen6-.5\wd1
                \kern\dimen6
                $\vcenter to\proofdotnumber\proofdotseparation
                        {\leaders\box1\vfill}$%
                \unhbox\proofrulename}%
\else   \dimen6=\fontdimen22\the\textfont2 
        \dimen7=\dimen6
        \advance\dimen6by.5\proofrulebreadth
        \advance\dimen7by-.5\proofrulebreadth
        \setbox0=\hbox{%
                \kern\shortenproofleft
                \ifdoubleproof
                \then   \hbox to\dimen0{%
                        $\mathsurround0pt\mathord=\mkern-6mu%
                        \cleaders\hbox{$\mkern-2mu=\mkern-2mu$}\hfill
                        \mkern-6mu\mathord=$}%
                \else   \vrule height\dimen6 depth-\dimen7 width\dimen0
                \fi
                \unhbox\proofrulename}%
        \ht0=\dimen6 \dp0=-\dimen7
\fi
\let\doll\relax
\ifwasinsideprooftree
\then   \let\VBOX\vbox
\else   \ifmmode\else$\let\doll=$\fi
        \let\VBOX\vcenter
\fi
\VBOX   {\baselineskip\proofrulebaseline \lineskip.2ex
        \expandafter\lineskiplimit\ifproofdots0ex\else-0.6ex\fi
        \hbox   spread\dimen5   {\hfi\unhbox\proofabove\hfi}%
        \hbox{\box0}%
        \hbox   {\kern\dimen2 \box\proofbelow}}\doll%
\global\dimen2=\dimen2
\global\dimen3=\dimen3
\egroup 
\ifonleftofproofrule
\then   \shortenproofleft=\dimen2
\fi
\shortenproofright=\dimen3
\onleftofproofrulefalse
\ifinsideprooftree
\then   \hskip.5em plus 1fil \penalty2
\fi
}


%% file: 1-Engineering.tex
\medskip\section{Old security paradigms}
\label{Old}
Security means many things to many people. For a 
software engineer, it often means that there are no buffer 
overflows or dangling pointers in the code. For a 
cryptographer, it means that any successful attack on the 
cypher can be reduced to an algorithm for computing discrete logarithms, or to integer factorization. For a diplomat, security means that the enemy  cannot read the confidential messages. For a credit card operator, it means that the total costs of the fraudulent transactions and of the measures to prevent them are low, 
relative to the revenue. For a bee, security means that no intruder into the beehive  will escape her sting\ldots

Is it an accident that all these different ideas 
go under the same name? What do they really have in 
common? They are studied in different sciences, ranging from computer science to biology, by a wide variety of different methods. Would it be useful to study them together?

%

\bigskip

\subsection{What is security?}
If all avatars of security have one thing in common, it is 
surely the idea that \emph{there are enemies and potential attackers out there}. All security concerns, from computation to politics and biology, come down to averting the adversarial processes in the environment, that are poised to subvert the goals of the system. There are, for instance, many kinds of bugs in software, but only those that the hackers use are a security concern.  

In all engineering disciplines, the system guarantees a 
functionality, provided that the environment satisfies some 
assumptions. This is the standard \emph{assume-guarantee} format of the engineering correctness statements. Such statements are useful when the environment is passive, so that the assumptions about it remain valid for a while. {The essence of security engineering is that the environment actively seeks to invalidate system's assumptions.} 

Security is thus an \emph{adversarial process}. In all 
engineering disciplines, failures usually arise from engineering errors and noncompliance. In security, failures arise \emph{in spite} of the compliance with the best engineering practices of the moment. Failures are the first class citizens of security: every key has a lifetime, and in a sense, every system too. For all major software systems, we normally expect security updates, which usually arise from attacks, and often inspire them. 


\medskip\subsection{Where did security come from?}
The earliest examples of security technologies are found 
among the earliest documents of civilization. Fig.~\ref
{bulla} shows security tokens with a tamper protection 
technology from almost 6000 years ago. Fig.\ref{MLsec} 
depicts the situation where this technology was 
probably used. Alice has a lamb and Bob has built a secure vault, perhaps with multiple security levels, spacious enough to store both Bob's and Alice's assets. For each of Alice's assets deposited in the vault, Bob issues a 
clay token, with an inscription identifying the asset. Alice's tokens are then encased into a \emph{bulla}, a round, hollow "envelope" of clay, which is then baked to prevent tampering. When she wants to withdraw her deposits, Alice submits her bulla to 
Bob, he breaks it, extracts the tokens, and returns the goods. Alice can also give her bulla to Carol, who can also submit it to Bob, to withdraw the goods, or pass on to Dave. Bull\ae\ can thus be traded, and they facilitate exchange 
economy. The tokens used in the bull\ae\ evolved into the earliest forms of money, and the inscriptions on them led to the earliest numeral systems, as well as to Sumerian cuneiform script, which was one of the earliest alphabets. Security thus predates literature, science, mathematics, and even money.
\begin{figure*}[htdp]
\begin{minipage}[b]{0.47\linewidth}
\centering
\includegraphics[height=5cm,clip=true,trim=0 16 0 0]
{PICS/clay-envelope.epsf}
\caption{Tamper protection from 3700 BC}
\label{bulla}
\end{minipage}
\hspace{.8cm}
\begin{minipage}[b]{0.4\linewidth}
\centering
\newcommand{\alice}{\mbox{\small Alice}}
\newcommand{\bob}{\mbox{\small Bob}}
\newcommand{\one}{}
\newcommand{\two}{}
\newcommand{\three}{}
\newcommand{\four}{}
\newcommand{\five}{}
\input{PICS/pic-vault}
\caption{To withdraw her sheep from Bob's secure vault, 
Alice submits a tamper-proof token from Fig.~\ref{bulla}.}
\label{MLsec}
\end{minipage}
\end{figure*}

\medskip\subsection{Where is security going?}
Through history, security technologies evolved gradually, serving the purposes of war and peace, protecting public resources and private property. As computers pervaded all aspects of social life, security became interlaced with computation, and security engineering came to be closely related with computer science. The developments in the realm of security are nowadays inseparable from the developments in the realm of computation. The most notable such development is, of course, \emph{cyber space}.

\medskip

\subsubsection*{Paradigms of computation}
In the beginning, engineers built computers, and wrote 
programs to control computations. The platform of computation was the computer,  and it was used to execute algorithms and calculations, allowing people to discover, e.g., fractals, and to invent 
compilers, that allowed them to write and execute more 
algorithms and more calculations more efficiently. Then the 
operating system became the platform of computation, and 
software was developed on top of it. The era of personal 
computing and enterprise software broke out. And then the 
Internet happened, followed by cellular networks, and 
wireless networks, and ad hoc networks, and mixed networks. {Cyber space emerged as the distance-free space of instant, costless communication.} 
Nowadays software is developed to run in cyberspace. The Web is, strictly speaking, just a software system, albeit a formidable one. A botnet is also a software system. As social space blends with cyber space, many social (business, collaborative) processes can be usefully construed as software systems, that ran on social networks as hardware. Many social and computational processes become inextricable. Table~\ref{ages-comp} summarizes the crude picture of the paradigm shifts which led to this remarkable situation.

\begin{table*}[htdp]
\begin{center}
\begin{tabular}{|m{2.5cm}||m{4cm}|m{4cm}|m{4cm}|m{4cm}|}
\hline
\textit{\textbf{age}} & \textit{ancient times} & \textit{middle ages} & \textit{modern times} \\[1ex]
\hline \hline
\textbf{platform}  & computer & operating system & network \\[1ex]
\hline
 \textbf{applications} & Quicksort, compilers & MS Word, Oracle  &   WWW, botnets \\[1ex]
\hline
\textbf{requirements} & correctness, termination & liveness, safety 
& trust, privacy \\[1ex]
 \hline
\textbf{tools} & 
programming languages & 
specification languages & 
 scripting languages \\[1ex]
 \hline
\end{tabular}
\end{center}
\caption{Paradigm shifts in computation}
\label{ages-comp}
\end{table*}%

But as every person got connected to a computer, and every computer to a  network, and every network to a network of networks,  computation became interlaced with communication, and ceased to be programmable.  The functioning of the Web and of web applications is not determined by the code in the same sense as in a traditional software system: after all, web applications do include the human users as a part of their runtime. The fusion of social and computational processes in cyber-social space leads to a new type of information processing, where the purposeful program executions at the network nodes are supplemented by spontaneous data-driven evolution of network links. While the network emerges as the new computer, data and metadata become inseparable, and a new type of security problems arises.

\medskip
\subsubsection*{Paradigms of security}
In early computer systems, security tasks mainly concerned sharing of the computing resources. In computer networks, security goals expanded to include information protection. Both computer security and information security essentially depend on a clear  distinction between the secure areas, and the insecure areas, separated by a security perimeter. Security engineering caters for computer security and for information security by providing the tools to build and defend the security perimeter. In cyber space, the secure areas are separated from the insecure areas by the "walls" of cryptography; and they are connected by the "gates" of cryptographic protocols.\footnote{This is, of course, a blatant oversimplification, as are many other statements I make. In a sense, every statement is an oversimplification of reality, abstracting away the matters deemed irrelevant. The gentle reader is invited to squint whenever any of the details that I omit do seem relevant, and add them to the picture. The shape of a forest should not change when some trees are enhanced.}  But as networks of computers and devices spread through physical and social spaces, the distinctions between the secure and the insecure areas become blurred. 
\begin{table*}[htdp]
\begin{center}
\begin{tabular}{|m{2.5cm}||m{4cm}|m{4cm}|m{4cm}|m{4cm}|}
\hline
\textit{\textbf{age}} &  \textit{middle ages} & \textit{modern times} & \textit{postmodern times} \\[1ex]
\hline \hline
\textbf{space}  & computer center & cyber space & cyber-social space \\[1ex]
\hline
 \textbf{assets} & computing resources  & 
information & public and private resources  \\[1ex]
\hline
\textbf{requirements} & availability, authorization & integrity, confidentiality 
& trust, privacy \\[1ex]
 \hline
\textbf{tools} &  locks, tokens, passwords & cryptography, protocols & mining and classification \\[1ex]
 \hline
\end{tabular}
\end{center}
\caption{Paradigm shifts in security}
\label{ages-sec}
\end{table*}%
And in such areas of cyber-social space, information processing does not yield to programming, and cannot be secured just by cryptography and protocols. What else is there?

%% file: PICS/pic-vault.tex
\ifx\JPicScale\undefined\def\JPicScale{1}\fi
\psset{unit=\JPicScale mm}
\psset{linewidth=0.3,dotsep=1,hatchwidth=0.3,hatchsep=1.5,shadowsize=1,dimen=middle}
\psset{dotsize=0.7 2.5,dotscale=1 1,fillcolor=black}
\psset{arrowsize=1 2,arrowlength=1,arrowinset=0.25,tbarsize=0.7 5,bracketlength=0.15,rbracketlength=0.15}
\begin{pspicture}(0,0)(55,40)
\newrgbcolor{userFillColour}{0.87 0.76 0.76}
\pspolygon[fillcolor=userFillColour,fillstyle=solid](0,40)(30,40)(30,20)(0,20)
\newrgbcolor{userFillColour}{0.84 0.84 0.64}
\pspolygon[fillcolor=userFillColour,fillstyle=solid](0,0)(10,0)(10,30)(0,30)
\newrgbcolor{userFillColour}{0.64 0.52 0.35}
\pspolygon[fillcolor=userFillColour,fillstyle=solid](0,20)(10,20)(10,30)(0,30)
\newrgbcolor{userFillColour}{0.73 0.84 0.87}
\pspolygon[fillcolor=userFillColour,fillstyle=solid](45,40)(55,40)(55,0)(45,0)
\newrgbcolor{userFillColour}{0.77 0.89 0.88}
\rput[br](29.38,0.62){$\one$}
\newrgbcolor{userFillColour}{0.77 0.89 0.88}
\rput[br](29.38,20.62){$\two$}
\newrgbcolor{userFillColour}{0.77 0.89 0.88}
\rput[bl](0.62,0.62){$\three$}
\newrgbcolor{userFillColour}{0.77 0.89 0.88}
\rput[br](54,1){$\four$}
\rput{0}(36.12,35.12){\psellipse[linewidth=0.15](0,0)(1,-1)}
\newrgbcolor{userFillColour}{0.77 0.89 0.88}
\psline[linewidth=0.15,fillcolor=userFillColour,fillstyle=solid](36.12,34.12)(36.12,31.12)
\newrgbcolor{userFillColour}{0.77 0.89 0.88}
\psline[linewidth=0.15,fillcolor=userFillColour,fillstyle=solid](36.12,31.12)(35.12,28.12)
\newrgbcolor{userFillColour}{0.77 0.89 0.88}
\psline[linewidth=0.15,fillcolor=userFillColour,fillstyle=solid](36.12,31.12)(37.12,28.12)
\newrgbcolor{userFillColour}{0.77 0.89 0.88}
\psline[linewidth=0.15,fillcolor=userFillColour,fillstyle=solid](36.12,33.12)(38.12,32.12)
\newrgbcolor{userFillColour}{0.77 0.89 0.88}
\psline[linewidth=0.15,fillcolor=userFillColour,fillstyle=solid](36.12,33.12)(34.12,32.12)
\rput{0}(21,35){\psellipse[linewidth=0.15](0,0)(1,-1)}
\newrgbcolor{userFillColour}{0.77 0.89 0.88}
\psline[linewidth=0.15,fillcolor=userFillColour,fillstyle=solid](21,34)(21,31)
\newrgbcolor{userFillColour}{0.77 0.89 0.88}
\psline[linewidth=0.15,fillcolor=userFillColour,fillstyle=solid](21,31)(20,28)
\newrgbcolor{userFillColour}{0.77 0.89 0.88}
\psline[linewidth=0.15,fillcolor=userFillColour,fillstyle=solid](21,31)(22,28)
\newrgbcolor{userFillColour}{0.77 0.89 0.88}
\psline[linewidth=0.15,fillcolor=userFillColour,fillstyle=solid](21,33)(23,32)
\newrgbcolor{userFillColour}{0.77 0.89 0.88}
\psline[linewidth=0.15,fillcolor=userFillColour,fillstyle=solid](21,33)(19,32)
\newrgbcolor{userFillColour}{0.77 0.89 0.88}
\newrgbcolor{userHatchColour}{1 1 0.8}
\rput{0}(6,34.75){\psellipse[linewidth=0.05,fillcolor=userFillColour,fillstyle=crosshatch*,hatchwidth=0.05,hatchsep=0.3,hatchcolor=userHatchColour](0,0)(4,-2.5)}
\newrgbcolor{userFillColour}{0.77 0.89 0.88}
\psline[fillcolor=userFillColour,fillstyle=solid](4.25,32.38)(4,31.24)
\newrgbcolor{userFillColour}{0.77 0.89 0.88}
\psline[fillcolor=userFillColour,fillstyle=solid](5,32.25)(5,31.24)
\newrgbcolor{userFillColour}{0.77 0.89 0.88}
\psline[fillcolor=userFillColour,fillstyle=solid](8,32.38)(8.62,31.12)
\newrgbcolor{userFillColour}{0.77 0.89 0.88}
\psline[fillcolor=userFillColour,fillstyle=solid](7.38,32.38)(7.38,31.12)
\newrgbcolor{userFillColour}{0.77 0.89 0.88}
\newrgbcolor{userHatchColour}{1 1 0.8}
\rput{90}(9.88,36.75){\psellipse[linewidth=0.05,fillcolor=userFillColour,fillstyle=crosshatch*,hatchwidth=0.05,hatchsep=0.3,hatchcolor=userHatchColour](0,0)(1.19,-1.12)}
\rput{112.29}(9.25,37.37){\psellipticarc[linewidth=0.2](0,0)(0.85,-0.58){-25.6}{109.12}}
\rput{111.93}(9.88,37.37){\psellipticarc[linewidth=0.2](0,0)(0.85,-0.59){-25.77}{108.95}}
\newrgbcolor{userFillColour}{0.77 0.89 0.88}
\rput[t](36.25,27.5){$\alice$}
\newrgbcolor{userFillColour}{0.77 0.89 0.88}
\rput(20.88,26.5){$\bob$}
\newrgbcolor{userFillColour}{0.77 0.89 0.88}
\rput[bl](0.62,20.62){$\five$}
\newrgbcolor{userFillColour}{1 0.8 0}
\rput{130.66}(5.42,24.82){\psellipticarc[linewidth=0.1,fillcolor=userFillColour,fillstyle=solid](0,0)(2,-1.93){68.28}{370.3}}
\newrgbcolor{userFillColour}{1 0.8 0}
\rput{77.85}(6.88,28.61){\psellipticarc[linewidth=0.1,fillcolor=userFillColour,fillstyle=solid](0,0)(2.12,-1.12){-171.14}{-75.42}}
\newrgbcolor{userFillColour}{1 0.8 0}
\rput{79.12}(3.82,27.72){\psellipticarc[linewidth=0.1,fillcolor=userFillColour,fillstyle=solid](0,0)(2.02,-1.02){25.73}{115.61}}
\end{pspicture}

%% file: 2-Science.tex
\bigskip\section{A second-hand but almost-new security paradigm: Know your enemy}\label{New}
\subsection{Security beyond architecture
}\label{Fortification}
\begin{figure}[htbp]
\begin{center}
\includegraphics[height=5.7cm
]{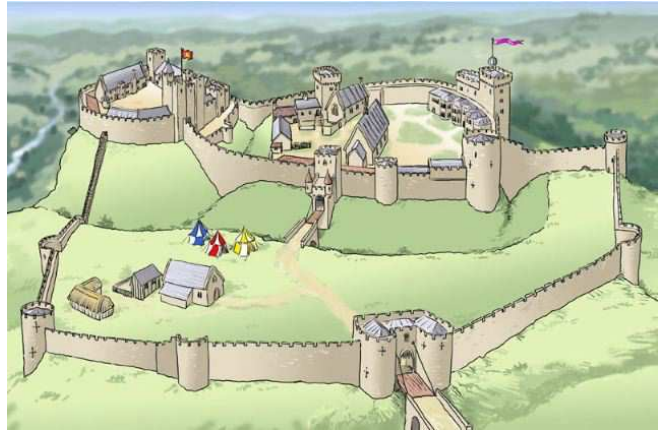}
\caption{Static security}
\label{MLS-castle}
\end{center}
\end{figure}

\begin{figure}[htbp]
\begin{center}
\includegraphics[height=5.7cm,clip=true,trim = 0 180 0 40]{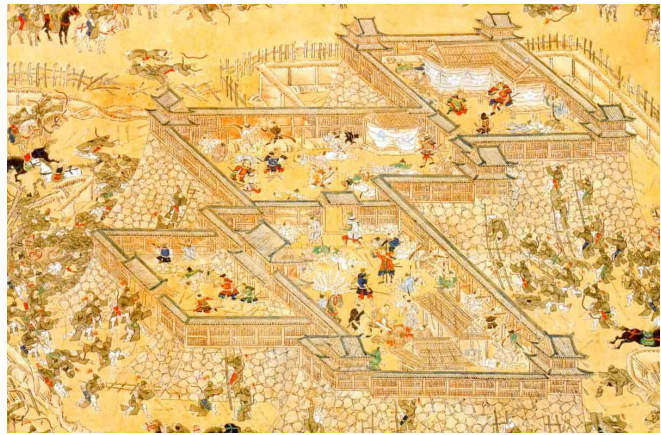}
\caption{Dynamic security}
\label{battle}
\end{center}
\end{figure}

Let us take a closer look at the paradigm shift to postmodern cyber security in Table \ref{ages-sec}. It can be illustrated as the shift from Fig.~\ref{MLS-castle} to Fig.~\ref{battle}. The fortification in Fig.~\ref{MLS-castle} represents the view that security is in essence an architectural task. A fortress consists of walls and gates, 
separating the secure area within from the insecure area 
outside. The boundary between these two areas is the 
security perimeter. The secure area may be further 
subdivided into the areas of higher security and the areas of lower security. In cyber space, as we mentioned,
the walls are realized using crypto systems, whereas the 
gates are authentication protocols.  
But as every fortress owner knows, the walls and  the 
gates are not enough for security: you also need some soldiers to defend it, and some weapons to arm the soldiers, and some craftsmen to build the weapons, and so on. Moreover, you also need police and judges to maintain security within the fortress. They take care for the dynamic aspects of security. These dynamic aspects arise from the fact that sooner or later, the enemies will emerge inside the fortress: they will scale the walls at night (i.e. break the crypto), or sneak past the gatekeepers (break the protocols), or build up trust and enter with honest intentions, and later defect to the enemy; or enter as moles, with the intention to strike later. In any case, security is not localized at the security perimeters of Fig.~\ref{MLS-castle}, but evolves in-depth, like on Fig.~\ref{battle}, through social processes, like  trust, privacy, reputation, influence.

In summary, besides the methods to keep the attackers out, security is also concerned with the methods to deal with the attackers once they get in. Security researchers have traditionally devoted more attention to the former family of methods. Insider threats have attracted a lot of attention recently, but a coherent set of research methods is yet to emerge.

Interestingly, though, there is a sense in which security becomes an easier task when the attacker is in. Although unintuitive at the first sight, this idea becomes natural when security processes are viewed in a broad context of the information flows surrounding them (and not only with respect to the data designated to be secret or private). To view security processes in this broad context, it is convenient to model them as \emph{games of incomplete information} \cite{Aumann-Heifetz:handbook-incomplete}, where the players do not have enough information to predict the opponent's behavior.  For the moment, let me just say that the two families of security methods (those to keep the attackers out, and those to catch them when they are in) correspond to two families of strategies in certain games of incomplete information, and turn out to have quite different winning odds for the attacker, and for defender. In fact, they have the \emph{opposite} winning odds. 

In the fortress mode, when the defenders' goal is to keep the attackers out, it is often observed that the attackers only need to find one attack vector to enter the fortress, whereas the defenders must defend all attack vectors to prevent them. When the battle switches to the dynamic mode, and the defense moves inside, then the defenders only need to find one marker to recognize and catch the attackers, whereas the attackers must cover all their markers. This strategic advantage is also the critical aspect of the immune response, where the invading organisms are purposely sampled and analyzed for chemical markers. Some aspects of this observation have, of course, been discussed within the framework of biologically inspired security. Game theoretic modeling seems to be opening up a new dimension in this problem space. We present a sketch to illustrate this new technical and conceptual direction.

\medskip

\subsection{The game of attack vectors}\label{Game}
\paragraph{Arena} Two players, the attacker $A$ and the defender $D$, battle for some assets of value to both of them. They are given equal, disjoint territories, with the borders of equal length, and equal amounts of force, expressed as two vector fields distributed along their respective borders. The players can redistribute the forces and move the borders of their territories. The territories can thus take any shapes and occupy any areas where the players may move them, obeying the constraints that
\begin{rnumerate}
\item the length of the borders of both territories must be preserved, and
\item the two territories must remain disjoint, except that they may touch at the borders.
\end{rnumerate}
It is assumed that the desired asset $\Theta$ is initially held by the defender $D$. Suppose that storing this asset takes an area of size $\theta$. Defender's goal is thus to maintain a territory $p_D$ with an area $\int p_D \geq \theta$.  Attacker's goal is to decrease the size of $p_D$ below $\theta$, so that the defender must release some of the asset $\Theta$. To achieve this, the attacker $A$ must bring his\footnote{I hope no one minds that I will be using \emph{"he"} for both $A$ttacker and $D$efender, in an attempt to avoid distracting connotations.} forces to defender $D$'s borders, and push into his territory. A position in the game can thus be something like Fig.~\ref{init}.  

\paragraph{Game} At each step in the game, each player makes a move by specifying a distribution of his forces along his borders. Both players are assumed to be able to redistribute their forces with equal agility. The new force vectors meet at the border, they add up, and the border moves along the resulting vector. So if the vectors are, say, in the opposite directions, the forces subtract and the border is pushed by the greater vector. 

The players observe each other's positions and moves in two ways:
\begin{anumerate}
\item Each player knows his own moves, i.e. distributions, and sees how his borders change. From the change in the previous move, he can thus derive the opponent's current distribution of the forces along the common part of the border.
\item Each player sees all movement in the areas enclosed within his territory, i.e. observes any point on a straight line between any two points that he controls. That means that each player sees the opponent's next move at all points that lie within the convex hull of his territory, which we call \emph{range}.
\end{anumerate}
According to (b), the position in Fig.~\ref{init} allows $A$ to see $D$'s next move. $D$, on the other hand only gets to know $A$'s move according to (a), when his own border changes. This depicts, albeit very crudely, the information asymmetry between the attacker and the defender. 

\paragraph{Question} \emph{How should rational players play this game?}

\medskip

\subsubsection{Fortification strategy} 
\paragraph{Goals} Since each player's total force is divided by the length of his borders, the maximal area defensible by a given force has the shape of a disk. All other shapes with the boundaries of the same length enclose smaller areas. So $D$'s simplest strategy is to acquire and maintain the smallest disk shaped territory of size $\theta$.\footnote{This is why the  core of a medieval fortification was a round tower with a thick wall and a small space inside. The fortress itself often is not round, because the environment is not flat, or because the straight walls were easier to build; but it is usually at least convex. Later fortresses, however, had protruding towers --- to attack the attacker. Which leads us beyond the fortification strategy\ldots} This is the \emph{fortification} strategy: $D$ only responds to $A$'s moves. 

$A$'s goal is, on the other hand, to create "dents" in $D$'s territory $p_D$, since the area of $p_D$ decreases most when its convexity is disturbed. If a dent grows deep enough to reach across $p_D$, or if two dents in it meet, then $p_D$ disconnects in two components. Given a constant length of the border, it is easy to see that the size of the enclosed area decreases exponentially as it gets broken up. In this way, the area enclosed by a  border of given length can be made arbitrarily small. 

But how can $A$ create dents? Wherever he pushes, the defender will push back. Since their forces are equal and constant, increasing the force along one vector decreases the force along another vector.

\begin{figure}
\begin{minipage}[b]{0.43\linewidth}
\centering
\def\JPicScale{.27}
\input{PICS/1-fortify.tex}
\caption{Fortification}
\label{init}
\end{minipage}
\hspace{.6cm}
\begin{minipage}[b]{0.43\linewidth}
\centering
\def\JPicScale{.27}
\input{PICS/2-honeypot.tex}
\caption{Honeypot}
\end{minipage}
\label{honeypot}
\end{figure}

\paragraph{Optimization tasks} To follow the fortification strategy, $D$ just keeps restoring $p_D$ to a disk of size $\theta$. To counter $D$'s defenses, $A$ needs to find out where they are the weakest. He can observe this wherever $D$'s territory $p_D$ is within $A$'s range, i.e. contained in the convex hull of $p_A$. So $A$ needs to maximize the intersection of his range with $D$'s territory. Fig.~\ref{init} depicts a position where this is achieved: $D$ is under $A$'s siege. It embodies the Fortification Principle, that the defender  must defend all attack vectors, whereas the attacker only needs to select one. For a fast push, $A$ randomly selects an attack vector, and waits for $D$ to push back. Strengthening $D$'s defense along one vector weakens it along another one. Since all of $D$'s territory is within $A$'s range, $A$ sees where $D$'s defense is the weakest, and launches the next attack there. In contrast, $D$'s range is initially limited to his own disk shaped territory. So $D$ only "feels" $A$'s pushes when his own borders move. At each step, $A$ pushes at $D$'s weakest point, and creates a deeper dent. $A$ does enter into $D$'s range, but $D$'s fortification strategy makes no use of the information that could be obtained about $A$. The number of steps needed to decrease $p_D$ below $\theta$ depends on how big are the forces and how small are the contested areas. 
%

\medskip
\subsubsection{Adaptation strategy}
What can $D$ do to avoid the unfavorable outcome of the fortification strategy? The idea is that he should learn to know his enemy: he should also try to shape his territory to maximize the intersection of his range with $A$'s territory. $D$ can lure $A$ into his range simply letting $A$ dent his territory.  This is the familiar \emph{honeypot} approach, illustrated on Fig.~6. Instead of racing around the border to push back against every attack, $D$ now gathers information about $A$'s next moves within his range. For this, he sacrifices a little bit of his territory, as a bait (the "honey") for $A$. If $A$ sticks with his strategic preferences, he will accept $D$'s sacrifice, and enter into $D$'s range more and more.
\begin{figure}
\begin{minipage}[b]{0.43\linewidth}
\centering
\def\JPicScale{.25}
\input{PICS/3-sampling.tex}
\caption{Sampling}
\label{sample}
\end{minipage}
\hspace{.6cm}
\begin{minipage}[b]{0.43\linewidth}
\centering
\def\JPicScale{.27}
\input{PICS/4-adapt.tex}
\caption{Adaptation}
\end{minipage}
\label{adapt}
\end{figure}
Fig.~7 depicts a further step in $D$'s strategic development, where he does not just passively wait for $A$ to enter the honeypot, but actively herds $A$ into it. This is the \emph{sampling} strategy. Formally, it is characterized by $D$'s higher valuation for the information about $A$, than for the territory alone. This is reflected in the fact that $D$'s territory gradually evolves into a shape optimized for information gathering. In the cyber wars of the day, the sampling strategy is emerging, e.g., among the researchers who have gone beyond luring some bots into some sandboxed computers, and hijacked parts of botnets from their owners, for the sole purpose of research \cite{CovaM:CCS09,CovaM:SP11}. Continuing in this direction leads to the long term strategy of \emph{adaptation}, depicted on  Fig.~8, where all of $D$'s strategic valuation is assigned to the information about the opponent.  Here $A$'s attacks are actively observed and prevented; the territory is maintained as a side effect of keeping the opponent localized. In the long term, $D$ wins. The proviso is that $D$ has enough territory to begin with.  A simplifying assumption of the presented model is that $A$ blindly sticks with his valuation of the territory, leading him to accept all baits. Reality is, of course, not so simple, and $A$'s strategy also allows various refinements. However, to achieve his goal of stealing $D$'s  assets, $A$ cannot avoid entering $D$'s range altogether. $D$, on the other hand, cannot allow that the size of his territory drops below $\theta$. Respecting these asymmetric constraints, both players' strategy refinements will evolve methods to trade territory for information, making increasingly efficient use of both. 
%
%
%
%
%

A formalism for a mathematical analysis of this game is sketched in the Appendix.

\medskip

\medskip\subsection{What does all this mean for security?}
The presented toy model provides a very crude picture of the evolution of defense strategies from fortification to adaptation. Intuitively, Fig.~5 can be viewed as a fortress under siege, whereas Fig.~8 can be interpreted as a macrophage localizing an invader. The intermediate pictures show the adaptive immune system luring the invader and sampling his chemical markers. 

But there is nothing exclusively biological about the adaptation strategy. Figures~5--8 could also be viewed entirely in the context of Figures~3--4, and interpreted as the transition from the medieval defense strategies to  modern political ideas. Fig.~8 could be viewed as a depiction of the idea of \emph{"preemptive siege"}: while the medieval rulers tried to keep their enemies out of their fortresses, some of the modern ones try to keep them in their jails. The evolution of strategic thinking illustrated on Figures~5-8 is pervasive in  all realms of security, i.e. wherever the adversarial behaviors are a problem, including cyber-security.

And although the paradigm of keeping an eye on your enemies is familiar, the fact that it reverts the odds of security and turns them in favor of the defenders does not seem to have received enough attention. It opens up a new game theoretic perspective on security, and suggests a new tool for it.

\bigskip

\section{A brand new security paradigm: Applied security by obscurity} \label{brand}
\subsection{Gaming security basics}\label{Ideas}
\paragraph{Games of information} In games of luck, each player has a \emph{type}, and some \emph{secrets}. The type determines player's preferences and behaviors. The secrets determine player's state. E.g., in poker, the secrets are the cards in player's hand, whereas her type consists of her risk aversion, her gaming habits etc.  \emph{Imperfect} information means that all players' types are  public information, whereas their states are unknown, because their secrets are private. In games of \emph{incomplete} information, both players' types and their secrets are unknown. The basic ideas and definitions of complete and incomplete information in games go all the way  back to von Neumann and Morgenstern \cite{VonNeumann-Morgenstern}. The ideas and techniques for modeling incomplete information are due to Harsanyi \cite{HarsanyiJ:bayesian}, and constitute an important part of game theory \cite{Mertens-Zamir,Fudenberg-Tirole:book,Aumann-Heifetz:handbook-incomplete}.

\paragraph{Security by secrecy} If cryptanalysis is viewed as a game, then the algorithms used in a crypto system can be viewed as the type of the corresponding player. The keys are, of course, its secrets. In this framewrok, Claude Shannon's slogan that \emph{"the enemy knows the system"} asserts that cryptanalysis should be viewed as a game of imperfect information. Since the type of the crypto system is known to the enemy, it is not a game of incomplete information. Another statement of the same imperative is the Kerckhoffs' slogan that \emph{"there is no security by obscurity"}. Here the obscurity refers to the type of the system, so the slogan thus suggests that the security of a crypto system should only depend on the secrecy of its keys, and remain secure if its type is known. In terms of physical security, both slogans thus say that the thief should not be able to get into the house without the right key, even if he knows the mechanics of the lock. The key is the secret, the lock is the type.

\paragraph{Security by obscurity} And while all seems clear, and we all pledge allegiance to Kerckhoffs's Principle, the practices of security by obscurity abound. E.g., besides the locks that keep the thieves out, many of us use some child-proof locks, to protect toddlers from dangers. A child-proof lock usually does not have a key, and only provides protection through the obscurity of its mechanism. 

On the cryptographic side, security by obscurity remains one of the main tools, e.g., in Digital Rights Management (DRM), where the task is to protect the digital content from its intended users. So our DVDs are encrypted to prevent copying; but the key must be on each DVD, or else the DVD could not be played. In order to break the copy protection, the attacker just needs to find out where to look for the key; i.e. he needs to know the system used to hide the key. For a sophisticated attacker, this is no problem; but the majority is not sophisticated. The DRM is thus based on the second-hand but almost-new paradigm from the preceding section: the DVD designers study the DVD users and hide the keys in obscure places. From time to time, the obscurity wears out, by an advance in reverse engineering, or by a lapse of defenders attention\footnote{The DVD Copy Scramble System (CSS) was originally reverse engineered to allow playing DVDs on Linux computers. This was possibly facilitated by an inadvertent disclosure from the DVD Copy Control Association (CAA). DVD CAA pursued the authors and distributors of the Linux DeCSS module through a series of court cases, until the case was dismissed in 2004 \cite{EFF}. Ironically, the cryptography used in DVD CSS has been so weak, in part due to the US export controls at the time of design, that any computer fast enough to play DVDs could find the key by brute force within 18 seconds \cite{Frank}. This easy cryptanalytic attack was published before DeCSS, but seemed too obscure for everyday use.}. Security is then restored by analyzing the enemy, and either introducing new features to stall the ripping software, or by dragging the software distributors to court. Security by obscurity is an ongoing process, just like all of security.

\medskip\subsection{Logical complexity}
\smallskip\noindent\textbf{What is the difference between keys and locks?} The conceptual problem with the Kerckhoffs Principle, as the requirement that security should be based on secret keys, and not on obscure algorithms, is that it seems inconsistent, at least at first sight, with the Von Neumann architecture of our computers, where programs are represented as data. In a computer, both a key and an algorithm are strings of bits. Why can I hide a key and cannot hide an algorithm? More generally, why can I hide data, and cannot hide programs? 

Technically, the answer boils down to the difference between data encryption and program obfuscation. The task of encryption is to transform a data representation in such a way that it can be recovered if and only if you have a key. The task of obfuscation is to transform a program representation so that the obfuscated program runs roughly the same as the original one, but that the original code (or some secrets built into it) cannot be recovered. Of course, the latter is harder, because encrypted data just need to be secret, whereas an obfuscated program needs to be secret \emph{and} and to run like the original program. In \cite{no-obfuscation}, it was shown that some programs must disclose the original code in order to perform the same function (\emph{and} they disclose it in a nontrivial way, i.e. not by simply printing out their own code). The theory here confirms the empiric evidence that reverse engineering is, on the average\footnote{However, the \emph{International Obfuscated C Code Contest} \cite{IOCCC} has generated some interesting and extremely amusing work.}, effective enough that you don't want to rely upon its hardness. So it is much easier to find out the lock mechanism, than to find the right key, even in the digital domain. When they say that there is  no security by obscurity, security practitioners thus usually mean that reverse engineering is easy, whereas cryptanalysis is hard, and provides more durable security guarantees.\footnote{It is interesting to note that one of the initial ideas for public key crypto system, suggested in \cite{Diffie-Hellman}, was to partially evaluated a symmetric encryption module over its key, and  to publish its obfuscation as a public encryption module.}

\smallskip\noindent\textbf{One-way programming?} Modern cryptography is based on one-way functions, which are easy to compute, but hard to invert. Secure systems are designed to use the easy direction of one-way functions, whereas the attacks must invert them, i.e. compute the hard direction. A high level view thus displays security as a process where the attackers program attack algorithms in response to the algorithms of the system that they attack, whereas the defenders program system algorithms in response to some attacks. The question is now whether systems can be designed in such a way to make the defenders' programming tasks easy, and the attackers' programming tasks hard. Can we lift the idea of one-way functions to one-way programming?

Let us take a closer look. How can we make attacker's task harder? Since obfuscation is hard and reverse engineering is easy, we assume that the system code is accessible to the attacker, and that the attack code is accessible to the defender. This assumption is supported by the current security practices, where the attacker communities reverse engineer their target systems, and the security researchers decompile malware. But even when the code is completely transparent, the attacker's task of uncovering vulnerabilities of the system remains nontrivial. And even with a trove of detected vulnerabilities, the attacker still needs to design an effective attack algorithm to exploit them. The hard part of an attacker's job is thus the \emph{logical} task to analyze the system, and to design and implement an attack algorithm. The \emph{logical complexity} of such tasks is different from the \emph{computational complexity} of the system and the attack algorithms. Indeed, a computationally easy algorithm may be logically hard to construct, even when it can be expressed by a relatively succinct program, like e.g. \cite{AgrawalM:PRIMES}; whereas an algorithm that requires a minor logical effort to construct may, of course, require a great computational effort to run.

The different roles of the computational and the logical complexities in security can perhaps be pondered on the following example. In modern cryptography, a system $C$ would be considered very secure if an attack algorithm $A_C$ on it would yield a proof that $P=NP$. But how would you feel about a crypto system $L$ such that an attack algorithm $A_L$ would yield a proof that $P\neq NP$? What is the difference between the reductions
\[A_C\ \Longrightarrow\ P=NP \quad \mbox{ and }\quad A_L\ \Longrightarrow\ P\neq NP\quad\mbox{?}\]
The security of the system $C$ is based on the \emph{computational} complexity of the $NP$ problems.  The security of the system $L$ is based on the \emph{logical} complexity of proving $P\neq NP$. Most computer scientists believe that $P\neq NP$ is true. If $P=NP$ is thus false, then no attack on the system $C$ can exist, whereas an attack on the system $L$ may very well be possible. So the security of the system $L$ may very well be based on the \emph{obscurity} of the proof of $P\neq NP$, which most likely exists, but is very hard to find.  The best minds of mankind have spent many years looking for this proof, but has not yet managed to find it. Yet an attack $A_L$, together with the security reduction $A_L\ \Longrightarrow\ P\neq NP$, would provide such a proof. This attack would probably be welcomed with admiration and gratitude. In fact, the system $L$ is secure enough to protect a bank account with a \$1,000,000, since proving (or disproving) $P\neq NP$ is worth a Clay Institute Millenium Prize of \$ 1,000,000. If an attacker takes your money from the bank account, he will leave you with a proof worth much more. So the logical complexity of the system $L$ provides enough obscurity for a significant amount of security!

\medskip\noindent\textbf{But what is logical complexity?} Computational complexity of a program tells how many computational steps (counting them in time, memory, state changes, etc.) the program takes  to transform its input into its output. Logical complexity is not concerned with the execution of the program, but with its logical construction. Intuitively, if computational complexity of a program counts the number of computational steps needed to execute it on an input of a given length, its logical complexity should count the number of computational steps needed to derive that program from some given programming knowledge. For instance, while the computational complexity of an attack on a crypto system is the the number of computational steps that it requires to extract some information about the plaintext from the cyphertext, the logical complexity of that attack is the number of logical steps needed to find that attack algorithm from a given description of the system algorithm.  In other words, the logical complexity of an attack on a crypto system is the computational complexity of the task of finding a counterexample for the security claim of that system.\footnote{Such claims are usually stated in the form: "For all probabilistic polynomial-time Turing machines that the attacker may use, his advantage is not greater than [some formula]". The logical complexity of the attack is the computational complexity of attacker's task of finding that attack.}

The idea is thus to define logical complexity of an algorithm $A$ as the computational complexity of the fastest construction algorithm $P_A$ that outputs $A$, given some algorithmic knowledge as the input. The problem in formalizing this is that the fastest program that outputs $A$ is the program \texttt{print A}, which does not need any input, since $A$ is hardwired in it. To assure that $A$ is constructed in a nontrivial way, we need to look for a constructor $P_A$ that is the fastest among the algorithms that can be implemented by programs significantly \emph{shorter} than $A$ itself. 

This brings us into the realm of algorithmic information theory \cite{Chaitin,LevinL:conservation,Vitanyi:book}, where similar concepts, with slight variations, have been proposed under a variety of different names. Bennett's \emph{logical depth} \cite{BennettC:depth} seems to be the the closest. In Bennett's original formulation, logical depth is defined as a complexity measure assigned to  data, or to observations, although Bennett's fascinating analyses also assign logical depth to genes and to organisms, as the time that it took them to evolve \cite{BennettC:Review}. The step from logical depth of data to logical complexity of algorithms boils down to the view of algorithms-as-programs-as-data, originating from G\"odel's enumeration of recursive functions as numbers, and from Kleene's recursive indices. Towards a brief, somewhat oversimplified, but hopefully not misleading account of these formalisms, consider the  partial algebraic theory with two sorts,
\begin{itemize}
\item $\NNn$, representing the data, say as numbers, and 
\item $\MMm$, representing a family of algorithms, viewed as partial functions $\NNn^\ast \rightharpoonup \NNn$, say those that can be realized by Turing machines\footnote{More precisely, they are realized as \emph{self-delimiting} Turing machines, which allows them to receive multiple arguments on the same tape \cite{Levin-Zvonkin}.},
\end{itemize}
given together with the mappings
\bea \label{encoding}
\xymatrix{
\MMm \ar@/^/[rr]^{\enc{-}}  && \NNn \ar@/^/[ll]^{\dec{-}}
}
\eea
which make them isomorphic, i.e.
\[
\dec{\enc M} = M  \qquad \mbox{and}\qquad \enc{\dec n} = n
\]
This means not just that every machine $M\in \MMm$ can be encoded as a number $\enc M\in \NNn$, but also that any number $n\in \NNn$ can be viewed as a program corresponding to the machine $\dec n \in \MMm$, and executed on data. In addition, there are also 
\begin{itemize}
\item a \emph{universal composition machine} $U\in \MMm$, such that for all $P, Q \in \MMm$ and all $n\in \NNn$
\bea\label{universal}
U(\enc P, \enc Q,  n) & = & P(Q(n))
\eea
holds whenever either side is defined,
\item a \emph{time counter} $T\in \MMm$, where $T(\enc M, n)$ denotes the number of steps that the machine $M$ takes before it halts on the input $n$, and
\item a \emph{length function} $\ell:\NNn \to \NNn$, assigning a length to each piece of data.
\end{itemize}
Towards formalizing the above idea of logical complexity, we now define the \emph{algorithmic distance} $\CCC(A,B)$ between the algorithms $A,B\in \MMm$ to be the length of the shortest program $p$ that inputs the code $\enc A$ and outputs the code $\enc B$:
\bea\label{alg-dist}
\CCC(A,B) & = & \bigwedge_{\dec p \left(\enc A\right) = \enc B} \ell(p)
\eea
The \emph{logical distance} $\DDD(A,B)$ is now the shortest time that it takes to compute $\enc B$ from $\enc A$ by one of the shortest programs: 
\bea\label{log-dist}
\DDD(A,B) & = & \bigwedge_{\substack{\dec p \left(\enc A\right) = \enc B \\ \ell(p) = \CCC(A,B)}} T\left( p, \enc A\right)
\eea

\noindent{\bf Remarks.} Algorithmic distance is based on the relativized version of Solomonoff's and Kolmogorov's definitions of complexity \cite{SolomonoffR:inference,Kolmogorov}. Logical distance is based on a relativized and simplified version of Bennett's logical depth \cite{BennettC:depth}. It is  simplified in the sense that it does not take into account the possibility that slightly longer programs may run significantly faster. To capture this, the developed definitions of logical depth are parametrized over the difference in length between the fastest and the shortest programs. In some applications, this is an important technical detail, assuring the stability of the definition. In the current presentation, which is mainly conceptual, it would just complicate the definition. A \emph{conceptual} detail which is also omitted is that the universal composition machine, and the programs in \eqref{alg-dist} and \eqref{log-dist} need to be homomorphisms with respect to certain logical operations on programs. This additional requirement seems essential for the envisioned applications of logical distance as a tool of security by obscurity. Nevertheless, such matters must be left for future work. The path towards program and specification frameworks that would take into account the logical distances of algorithms, and advance the idea of one-programming, requires examining the diverse refinements of the basic idea of the  G\"odel-Kleene program encodings that in the meantime emerged from the theory and the extensive experience of program development \cite{Hoare:ProgPred,BorgerE:book-ASM,FiadeiroJ:book,PavlovicD:FOPS,Mossakowski:RelatingCASL,PavlovicD:SDR}. 

\smallskip\noindent\textbf{Logical security.} The idea of logical security is to make the derivation of an attack algorithm from a system algorithm logically complex. Formally, a system $S$ would thus be logically secure if the distance $\DDD(S, A)$ is large for all attacks $A$ on $S$. On the other hand, since the distance is obviously not symmetric, $\DDD(A,S)$ may be small, meaning that it may be easy to derive an improved system algorithm $S$ from an undesired algorithm $A$. This connects logical security with the idea of one-way programming.

At this point, the reader may object that it does not make much sense to call $\DDD$ a distance when $\DDD(A,S)$ is generally different from $\DDD(S,A)$. A possible answer to appeal to spatial intuition that $S$ lies lower than $A$, so that climbing from $S$ to $A$ requires energy, whereas descending from $A$ to $S$ releases it. But it gets worse. While algorithmic distance satisfies the triangle law 
\bea
\CCC(P,Q) + \CCC(Q,R) & \geq & \CCC(P,R)
\eea
realized by the universal composition machine, logical distance generally does not satisfy this law, since there may be a short but slow algorithm to construct $R$ from $P$, but it does not have to go through $Q$, and all algorithms to construct $Q$ from $P$ and $R$ from $Q$ may be long but fast. Nevertheless, logical distance can be easily shown to satisfy the law
\bea\label{triangle}
\DDD(P,Q)+\DDD(P\wedge Q, R) & \geq & \DDD(P,R\wedge Q) 
\eea
where $P\wedge Q$ denotes the parallel composition of $P$ and $Q$, i.e. an algorithm that satisfies $P\wedge Q(x,y) = (u,v)$ if and only if $P(x) = u$ and $Q(y) = v$, for all $x,y,u,v$. One is tempted to call this law \emph{"subbayesian"}, since it echoes the Bayes' law of conditional probabilities in the form
\bear
\Pr(r \wedge q \ |\  p) & = & \Pr(r\ |\  q\wedge p) \cdot \Pr(q\ |\ p)
\eear
where $p,q,r$ denote events. At any rate, we can now use \eqref{triangle} to bound the logical complexity $\DDD(S,A)$ of constructing an attack $A$ on a system $S$. For the particular case of the system $L$, given with a security reduction $A_L\Longrightarrow P\neq NP$, we first of all assume that we are given an effective algorithm\footnote{The \emph{constructivist} imperative that an implication like $A_L\Longrightarrow P\neq NP$ should be supported by an effective algorithm transforming any proof of the antecedent into a proof of the consequence was goes back to Brouwer \cite{TVD88}. Although constructivism was deemed impractical by most mathematicians, functional programming can be viewed as its practical realization \cite{Curry-Howard-iso,PavlovicD:constructions,PavlovicD:mapsII}. } to transform any attack $A_L$ to a proof of $P\neq NP$. Denote the time complexity of the shortest such algorithm by 
\bear
d & = & \DDD(A_L,\ P\neq NP)
\eear
where we abuse notation and write $P\neq NP$ for the algorithm that outputs a proof of the statement $P\neq NP$. Now if the system $L$ itself does not allow shortening of the proof $A_L \Longrightarrow P\neq NP$, i.e. if $d\geq \DDD(L\wedge A_L,\ P\neq NP$, then we can get
\bear
\DDD\left(L,\ A_L\right)\  + \ d\  & \geq &\   \DDD\left(L,\ P\neq NP\right)
\eear
as a substitution instance of \eqref{triangle}. Going back to security by obscurity, this means that, although  $L$ may be vulnerable to a computationally easy attack $A_L$, constructing this attack may be logically hard, nearly as hard as deriving a proof of $P\neq NP$ from it. 

\medskip
\paragraph{Logical complexity of randomized algorithms} While logical complexity as a tool and resource for security will clearly be built upon the deep and interesting results of algorithmic information theory,  it should be noted that a realistic model of attacker's logical practices will require algorithmic information theory of \emph{randomized} computation, since the algorithms used in security tend to be randomized. The encoding in \eqref{encoding} will thus not be strict but only approximate, with \eqref{universal} replaced with
\bea
\Pr\left( U(\enc P_\varepsilon , \enc Q_\varepsilon , x) = P(Q(x)) \right) & \gt & \varepsilon
\eea

\medskip

\subsection{Logical complexity of gaming security}
Randomized logical complexity brings us back to security as a game of incomplete information. In order to construct a strategy each player in such a game must supplement the available informations about the opponent by some \emph{beliefs}.  Mathematically, these beliefs have been modeled, ever since \cite{HarsanyiJ:bayesian}, as probability distributions over opponent's payoff functions. More generally, in games not driven by payoffs, beliefs can be modeled as probability distributions over the possible opponent's behaviors, or algorithms. In terms of randomized logical complexity, such a belief can thus be viewed as an approximate logical specification of opponent's algorithms. 

Since both players in a game of incomplete information are building beliefs about each other, they must also build beliefs about each other beliefs: $A$ formulates a belief about $B$'s belief about $A$, and $B$ formulates a belief about $A$'s belief about $B$. And so on to infinity. This is described in more detail in the Appendix, and in still more detail in \cite{Aumann-Heifetz:handbook-incomplete}. These hierarchies of beliefs are formalized as probability distributions over probability distributions. The framework of approximate logical complexity now captures the way in which the players encode their beliefs about each other's beliefs. This leads into an \emph{algorithmic} theory of incomplete information, where players' belief hierarchies consist of \emph{samplable} probability distributions.

%% file: PICS/1-fortify.tex
\ifx\JPicScale\undefined\def\JPicScale{1}\fi
\psset{unit=\JPicScale mm}
\psset{linewidth=0.3,dotsep=1,hatchwidth=0.3,hatchsep=1.5,shadowsize=1,dimen=middle}
\psset{dotsize=0.7 2.5,dotscale=1 1,fillcolor=black}
\psset{arrowsize=1 2,arrowlength=1,arrowinset=0.25,tbarsize=0.7 5,bracketlength=0.15,rbracketlength=0.15}
\begin{pspicture}(0,0)(139,125)
\newrgbcolor{userFillColour}{0.8 1 1}
\rput{0}(70,65){\psellipse[linewidth=1.15,fillcolor=userFillColour,fillstyle=solid](0,0)(50,-50)}
\newrgbcolor{userFillColour}{0.4 0 0}
\pscustom[linewidth=1.15,fillcolor=userFillColour,fillstyle=solid]{\psbezier(88,18.5)(91.5,19.55)(94.95,21.5)(99.5,25)
\psbezier(104.05,28.5)(107.35,31.8)(110.5,36)
\psbezier(113.65,40.2)(115.6,44.1)(117,49)
\psbezier(118.4,53.9)(119.15,57.65)(119.5,61.5)
\psbezier(119.85,65.35)(119.7,69.25)(119,74.5)
\psbezier(118.3,79.75)(116.35,84.55)(112.5,90.5)
\psbezier(108.65,96.45)(105.95,99.9)(103.5,102)
\psbezier(101.05,104.1)(97.9,106.05)(93,108.5)
\psbezier(88.1,110.95)(84.8,112.6)(82,114)
\psbezier(79.2,115.4)(77.1,117.2)(75,120)
\psbezier(72.9,122.8)(78.15,123.85)(92.5,123.5)
\psbezier(106.85,123.15)(116.45,117.9)(124.5,106)
\psbezier(132.55,94.1)(136.45,86.45)(137.5,80.5)
\psbezier(138.55,74.55)(139,68.7)(139,61)
\psbezier(139,53.3)(137.95,47)(135.5,40)
\psbezier(133.05,33)(130.05,27.75)(125.5,22.5)
\psbezier(120.95,17.25)(116.6,13.05)(111,8.5)
\psbezier(105.4,3.95)(99.7,0.95)(92,-1.5)
\psbezier(84.3,-3.95)(77.25,-5)(68.5,-5)
\psbezier(59.75,-5)(52.55,-3.5)(44.5,0)
\psbezier(36.45,3.5)(30.45,7.25)(24.5,12.5)
\psbezier(18.55,17.75)(14.5,22.4)(11,28)
\psbezier(7.5,33.6)(5.25,39)(3.5,46)
\psbezier(1.75,53)(1,59)(1,66)
\psbezier(1,73)(1.6,78.7)(3,85)
\psbezier(4.4,91.3)(6.95,96.7)(11.5,103)
\psbezier(16.05,109.3)(21.9,113.8)(31,118)
\psbezier(40.1,122.2)(48.2,124.15)(58,124.5)
\psbezier(67.8,124.85)(71.25,123.65)(69.5,120.5)
\psbezier(67.75,117.35)(64.6,115.4)(59,114)
\psbezier(53.4,112.6)(48,109.9)(41,105)
\psbezier(34,100.1)(29.5,95.3)(26,89)
\psbezier(22.5,82.7)(20.85,76.85)(20.5,69.5)
\psbezier(20.15,62.15)(20.75,56.3)(22.5,50)
\psbezier(24.25,43.7)(26.35,39.2)(29.5,35)
\psbezier(32.65,30.8)(35.65,27.8)(39.5,25)
\psbezier(43.35,22.2)(46.65,20.4)(50.5,19)
\psbezier(54.35,17.6)(57.05,16.7)(59.5,16)
\psbezier(61.95,15.3)(64.35,15)(67.5,15)
\psbezier(70.65,15)(73.65,15.3)(77.5,16)
\psbezier(81.35,16.7)(84.5,17.45)(88,18.5)
\closepath}
\rput(70,65){\System}
\rput(72,5){\Attack}
\psline[linewidth=0.5,arrowsize=2 2,arrowlength=1.5]{->}(43,89)(35,97)
\psline[linewidth=0.5,arrowsize=2 2,arrowlength=1.5]{->}(97,41)(105,33)
\psline[linewidth=0.5,arrowsize=2 2,arrowlength=1.5]{->}(46,39)(38,31)
\psline[linewidth=0.5,arrowsize=2 2,arrowlength=1.5]{->}(72,27)(72,18)
\psline[linewidth=0.5,arrowsize=2 2,arrowlength=1.5]{->}(96,90)(105,97)
\psline[linewidth=0.5,arrowsize=2 2,arrowlength=1.5]{->}(62,105)(60,111)
\psline[linewidth=0.5,arrowsize=2 2,arrowlength=1.5]{->}(107,65)(117,65)
\psline[linewidth=0.5,arrowsize=2 2,arrowlength=1.5]{->}(34,64)(23,64)
\psline[linewidth=0.5,arrowsize=2 2,arrowlength=1.5]{->}(81,104)(84,110)
\psline[linewidth=0.5,linecolor=white,arrowsize=2 2,arrowlength=1.5]{->}(129,42)(119,46)
\psline[linewidth=0.5,linecolor=white,arrowsize=2 2,arrowlength=1.5]{->}(11,90)(22,85)
\psline[linewidth=0.5,linecolor=white,arrowsize=2 2,arrowlength=1.5]{->}(40,7)(47,17)
\psline[linewidth=0.5,linecolor=white,arrowsize=2 2,arrowlength=1.5]{->}(129,87)(118,83)
\psline[linewidth=0.5,linecolor=white,arrowsize=2 2,arrowlength=1.5]{->}(46,119)(50,113)
\psline[linewidth=0.5,linecolor=white,arrowsize=2 2,arrowlength=1.5]{->}(10,38)(22,44)
\psline[linewidth=0.5,linecolor=white,arrowsize=2 2,arrowlength=1.5]{->}(99,10)(94,18)
\psline[linewidth=0.5,linecolor=white,arrowsize=2 2,arrowlength=1.5]{->}(101,116)(96,109)
\end{pspicture}

%% file: PICS/2-honeypot.tex
\ifx\JPicScale\undefined\def\JPicScale{1}\fi
\psset{unit=\JPicScale mm}
\psset{linewidth=0.3,dotsep=1,hatchwidth=0.3,hatchsep=1.5,shadowsize=1,dimen=middle}
\psset{dotsize=0.7 2.5,dotscale=1 1,fillcolor=black}
\psset{arrowsize=1 2,arrowlength=1,arrowinset=0.25,tbarsize=0.7 5,bracketlength=0.15,rbracketlength=0.15}
\begin{pspicture}(0,0)(148,118)
\newrgbcolor{userFillColour}{0.4 0 0}
\pscustom[linewidth=1.15,fillcolor=userFillColour,fillstyle=solid]{\psbezier(86.5,41.5)(87.55,36.95)(88.6,33.5)(90,30)
\psbezier(91.4,26.5)(94.55,26.2)(100.5,29)
\psbezier(106.45,31.8)(110.65,34.8)(114.5,39)
\psbezier(118.35,43.2)(120.75,47.55)(122.5,53.5)
\psbezier(124.25,59.45)(125.15,64.1)(125.5,69)
\psbezier(125.85,73.9)(125.4,78.1)(124,83)
\psbezier(122.6,87.9)(121.25,91.2)(119.5,94)
\psbezier(117.75,96.8)(115.95,99.65)(113.5,103.5)
\psbezier(111.05,107.35)(110.75,110.35)(112.5,113.5)
\psbezier(114.25,116.65)(117.4,117.85)(123,117.5)
\psbezier(128.6,117.15)(132.5,114.6)(136,109)
\psbezier(139.5,103.4)(142.05,97.85)(144.5,90.5)
\psbezier(146.95,83.15)(148,77.15)(148,70.5)
\psbezier(148,63.85)(147.7,58.6)(147,53)
\psbezier(146.3,47.4)(145.25,42.75)(143.5,37.5)
\psbezier(141.75,32.25)(139.5,27.75)(136,22.5)
\psbezier(132.5,17.25)(128.75,13.5)(123.5,10)
\psbezier(118.25,6.5)(111.65,4.25)(101.5,2.5)
\psbezier(91.35,0.75)(82.35,0)(71.5,0)
\psbezier(60.65,0)(53,0.75)(46,2.5)
\psbezier(39,4.25)(33,7.25)(26,12.5)
\psbezier(19,17.75)(14.5,22.25)(11,27.5)
\psbezier(7.5,32.75)(5.55,37.1)(4.5,42)
\psbezier(3.45,46.9)(2.55,51.4)(1.5,57)
\psbezier(0.45,62.6)(0.45,67.25)(1.5,72.5)
\psbezier(2.55,77.75)(3.75,82.25)(5.5,87.5)
\psbezier(7.25,92.75)(9.5,97.25)(13,102.5)
\psbezier(16.5,107.75)(20.55,110.9)(26.5,113)
\psbezier(32.45,115.1)(35.75,114.65)(37.5,111.5)
\psbezier(39.25,108.35)(38.8,105.05)(36,100.5)
\psbezier(33.2,95.95)(31.4,92.65)(30,89.5)
\psbezier(28.6,86.35)(27.7,83.5)(27,80)
\psbezier(26.3,76.5)(26.3,72.3)(27,66)
\psbezier(27.7,59.7)(29.05,54.45)(31.5,48.5)
\psbezier(33.95,42.55)(36.5,38.65)(40,35.5)
\psbezier(43.5,32.35)(47.25,29.95)(52.5,27.5)
\psbezier(57.75,25.05)(60.75,25.65)(62.5,29.5)
\psbezier(64.25,33.35)(65.6,36.8)(67,41)
\psbezier(68.4,45.2)(70.2,47.75)(73,49.5)
\psbezier(75.8,51.25)(78.2,51.4)(81,50)
\psbezier(83.8,48.6)(85.45,46.05)(86.5,41.5)
\closepath}
\newrgbcolor{userFillColour}{0.8 1 1}
\pscustom[linewidth=1.2,fillcolor=userFillColour,fillstyle=solid]{\psbezier(29,85.5)(27.6,80.95)(27,76.75)(27,71.5)
\psbezier(27,66.25)(27.6,61.6)(29,56)
\psbezier(30.4,50.4)(32.35,46.05)(35.5,41.5)
\psbezier(38.65,36.95)(43,33.35)(50,29.5)
\psbezier(57,25.65)(60.3,26.55)(61,32.5)
\psbezier(61.7,38.45)(62.6,42.35)(64,45.5)
\psbezier(65.4,48.65)(67.65,50.6)(71.5,52)
\psbezier(75.35,53.4)(78.65,53.55)(82.5,52.5)
\psbezier(86.35,51.45)(88.6,49.65)(90,46.5)
\psbezier(91.4,43.35)(92,39.3)(92,33)
\psbezier(92,26.7)(95.15,25.8)(102.5,30)
\psbezier(109.85,34.2)(114.2,37.8)(117,42)
\psbezier(119.8,46.2)(121.75,50.7)(123.5,57)
\psbezier(125.25,63.3)(125.85,67.95)(125.5,72.5)
\psbezier(125.15,77.05)(124.25,81.4)(122.5,87)
\psbezier(120.75,92.6)(118.8,96.05)(116,98.5)
\psbezier(113.2,100.95)(110.35,102.9)(106.5,105)
\psbezier(102.65,107.1)(97.25,108.3)(88.5,109)
\psbezier(79.75,109.7)(73.3,109.7)(67,109)
\psbezier(60.7,108.3)(56.2,107.55)(52,106.5)
\psbezier(47.8,105.45)(44.8,104.1)(42,102)
\psbezier(39.2,99.9)(36.95,97.95)(34.5,95.5)
\psbezier(32.05,93.05)(30.4,90.05)(29,85.5)
\closepath}
\rput(76,78){\System}
\rput(74,15){\Attack}
\psline[linewidth=0.5,arrowsize=2 2,arrowlength=1.5]{->}(46,83)(31,87)
\psline[linewidth=0.5,arrowsize=2 2,arrowlength=1.5]{->}(98,40)(106,32)
\psline[linewidth=0.5,arrowsize=2 2,arrowlength=1.5]{->}(55,41)(47,33)
\psline[linewidth=0.5,arrowsize=2 2,arrowlength=1.5]{->}(105,85)(119,86)
\psline[linewidth=0.5,arrowsize=2 2,arrowlength=1.5]{->}(67,54)(65,60)
\psline[linewidth=0.5,arrowsize=2 2,arrowlength=1.5]{->}(108,60)(119,57)
\psline[linewidth=0.5,arrowsize=2 2,arrowlength=1.5]{->}(43,59)(32,56)
\psline[linewidth=0.5,arrowsize=2 2,arrowlength=1.5]{->}(87,53)(90,59)
\psline[linewidth=0.5,linecolor=white,arrowsize=2 2,arrowlength=1.5]{->}(131,41)(121,45)
\psline[linewidth=0.5,linecolor=white,arrowsize=2 2,arrowlength=1.5]{->}(11,74)(24,74)
\psline[linewidth=0.5,linecolor=white,arrowsize=2 2,arrowlength=1.5]{->}(57,12)(56,24)
\psline[linewidth=0.5,linecolor=white,arrowsize=2 2,arrowlength=1.5]{->}(139,76)(127,76)
\psline[linewidth=0.5,linecolor=white,arrowsize=2 2,arrowlength=1.5]{->}(26,104)(33,98)
\psline[linewidth=0.5,linecolor=white,arrowsize=2 2,arrowlength=1.5]{->}(21,36)(33,42)
\psline[linewidth=0.5,linecolor=white,arrowsize=2 2,arrowlength=1.5]{->}(94,13)(95,24)
\psline[linewidth=0.5,linecolor=white,arrowsize=2 2,arrowlength=1.5]{->}(128,104)(120,98)
\psline[linewidth=0.5,arrowsize=2 2,arrowlength=1.5]{->}(89,100)(91,106)
\psline[linewidth=0.5,arrowsize=2 2,arrowlength=1.5]{->}(64,100)(62,106)
\psline[linewidth=0.5,linecolor=white,arrowsize=2 2,arrowlength=1.5]{->}(77,38)(77,50)
\psline[linewidth=0.5,linecolor=white,arrowsize=2 2,arrowlength=1.5]{->}(79,32)(87,35)
\psline[linewidth=0.5,linecolor=white,arrowsize=2 2,arrowlength=1.5]{->}(74,32)(65,35)
\end{pspicture}

%% file: PICS/3-sampling.tex
\ifx\JPicScale\undefined\def\JPicScale{1}\fi
\psset{unit=\JPicScale mm}
\psset{linewidth=0.3,dotsep=1,hatchwidth=0.3,hatchsep=1.5,shadowsize=1,dimen=middle}
\psset{dotsize=0.7 2.5,dotscale=1 1,fillcolor=black}
\psset{arrowsize=1 2,arrowlength=1,arrowinset=0.25,tbarsize=0.7 5,bracketlength=0.15,rbracketlength=0.15}
\begin{pspicture}(0,0)(185,132)
\newrgbcolor{userFillColour}{0.4 0 0}
\pscustom[linewidth=1.15,fillcolor=userFillColour,fillstyle=solid]{\psbezier(115,72.5)(112.9,64.45)(111.1,58.9)(109,54)
\psbezier(106.9,49.1)(105.85,44.3)(105.5,38)
\psbezier(105.15,31.7)(106.35,29.45)(109.5,30.5)
\psbezier(112.65,31.55)(115.35,32.75)(118.5,34.5)
\psbezier(121.65,36.25)(123.9,37.9)(126,40)
\psbezier(128.1,42.1)(129.9,44.65)(132,48.5)
\psbezier(134.1,52.35)(135.3,55.65)(136,59.5)
\psbezier(136.7,63.35)(137.75,66.95)(139.5,71.5)
\psbezier(141.25,76.05)(144.25,78.6)(149.5,80)
\psbezier(154.75,81.4)(158.5,80.05)(162,75.5)
\psbezier(165.5,70.95)(167.15,67.5)(167.5,64)
\psbezier(167.85,60.5)(167.55,56.6)(166.5,51)
\psbezier(165.45,45.4)(164.4,41.5)(163,38)
\psbezier(161.6,34.5)(160.25,31.8)(158.5,29)
\psbezier(156.75,26.2)(154.5,23.5)(151,20)
\psbezier(147.5,16.5)(144.35,13.95)(140.5,11.5)
\psbezier(136.65,9.05)(132.9,7.1)(128,5)
\psbezier(123.1,2.9)(117.85,1.7)(110.5,1)
\psbezier(103.15,0.3)(95.5,0)(85,0)
\psbezier(74.5,0)(67.75,0.45)(62.5,1.5)
\psbezier(57.25,2.55)(53.05,3.9)(48.5,6)
\psbezier(43.95,8.1)(40.05,10.8)(35.5,15)
\psbezier(30.95,19.2)(27.8,22.8)(25,27)
\psbezier(22.2,31.2)(20.4,34.05)(19,36.5)
\psbezier(17.6,38.95)(16.25,41.2)(14.5,44)
\psbezier(12.75,46.8)(11.7,49.2)(11,52)
\psbezier(10.3,54.8)(10.6,58.4)(12,64)
\psbezier(13.4,69.6)(15.2,73.05)(18,75.5)
\psbezier(20.8,77.95)(22.9,79.3)(25,80)
\psbezier(27.1,80.7)(29.2,80.55)(32,79.5)
\psbezier(34.8,78.45)(36.45,76.5)(37.5,73)
\psbezier(38.55,69.5)(39.6,66.35)(41,62.5)
\psbezier(42.4,58.65)(44.05,54.6)(46.5,49)
\psbezier(48.95,43.4)(52.25,39.5)(57.5,36)
\psbezier(62.75,32.5)(66.2,30.85)(69,30.5)
\psbezier(71.8,30.15)(73.3,31.8)(74,36)
\psbezier(74.7,40.2)(73.95,44.85)(71.5,51.5)
\psbezier(69.05,58.15)(66.95,64.3)(64.5,72)
\psbezier(62.05,79.7)(65.2,85.4)(75,91)
\psbezier(84.8,96.6)(93.35,96.75)(103.5,91.5)
\psbezier(113.65,86.25)(117.1,80.55)(115,72.5)
\closepath}
\newrgbcolor{userFillColour}{0.8 1 1}
\pscustom[linewidth=1.2,fillcolor=userFillColour,fillstyle=solid]{\psbezier(16.5,78.5)(21.75,83.75)(26.1,84.65)(31,81.5)
\psbezier(35.9,78.35)(38.6,73.85)(40,66.5)
\psbezier(41.4,59.15)(43.65,53.3)(47.5,47)
\psbezier(51.35,40.7)(56,36.35)(63,32.5)
\psbezier(70,28.65)(73.3,29.55)(74,35.5)
\psbezier(74.7,41.45)(71.85,50)(64.5,64)
\psbezier(57.15,78)(59.25,86.85)(71.5,93.5)
\psbezier(83.75,100.15)(94.55,100.15)(107.5,93.5)
\psbezier(120.45,86.85)(122.85,78.15)(115.5,64.5)
\psbezier(108.15,50.85)(105,42.3)(105,36)
\psbezier(105,29.7)(108.15,28.8)(115.5,33)
\psbezier(122.85,37.2)(127.2,40.8)(130,45)
\psbezier(132.8,49.2)(134.9,54.9)(137,64)
\psbezier(139.1,73.1)(142.7,78.5)(149,82)
\psbezier(155.3,85.5)(159.5,84.6)(163,79)
\psbezier(166.5,73.4)(170.55,72.35)(176.5,75.5)
\psbezier(182.45,78.65)(183.65,84.95)(180.5,96.5)
\psbezier(177.35,108.05)(168.35,115.7)(150.5,122)
\psbezier(132.65,128.3)(114.65,131.15)(90.5,131.5)
\psbezier(66.35,131.85)(49.1,129.45)(33,123.5)
\psbezier(16.9,117.55)(8.05,110.05)(3.5,98.5)
\psbezier(-1.05,86.95)(-1.2,80.35)(3,76.5)
\psbezier(7.2,72.65)(11.25,73.25)(16.5,78.5)
\closepath}
\psline[linewidth=0.5,linecolor=white,arrowsize=2 2,arrowlength=1.5]{->}(151,66)(140,70)
\psline[linewidth=0.5,arrowsize=2 2,arrowlength=1.5]{->}(25,99)(25,85)
\psline[linewidth=0.5,linecolor=white,arrowsize=2 2,arrowlength=1.5]{->}(123,23)(119,33)
\psline[linewidth=0.5,arrowsize=2 2,arrowlength=1.5]{->}(153,80)(142,76)
\psline[linewidth=0.5,arrowsize=2 2,arrowlength=1.5]{->}(91,102)(91,111)
\psline[linewidth=0.5,linecolor=white,arrowsize=2 2,arrowlength=1.5]{->}(24,64)(36,69)
\psline[linewidth=0.5,linecolor=white,arrowsize=2 2,arrowlength=1.5]{->}(86,84)(84,92)
\psline[linewidth=0.5,arrowsize=2 2,arrowlength=1.5]{->}(115,92)(119,97)
\psline[linewidth=0.5,arrowsize=2 2,arrowlength=1.5]{->}(105,98)(108,105)
\psline[linewidth=0.5,arrowsize=2 2,arrowlength=1.5]{->}(74,99)(70,105)
\psline[linewidth=0.5,arrowsize=2 2,arrowlength=1.5]{->}(156,101)(156,87)
\psline[linewidth=0.5,arrowsize=2 2,arrowlength=1.5]{->}(61,89)(56,94)
\psline[linewidth=0.5,arrowsize=2 2,arrowlength=1.5]{->}(55,51)(50,49)
\psline[linewidth=0.5,arrowsize=2 2,arrowlength=1.5]{->}(54,69)(44,65)
\psline[linewidth=0.5,arrowsize=2 2,arrowlength=1.5]{->}(50,85)(38,79)
\psline[linewidth=0.5,arrowsize=2 2,arrowlength=1.5]{->}(16,97)(6,79)
\psline[linewidth=0.5,arrowsize=2 2,arrowlength=1.5]{->}(164,97)(174,80)
\psline[linewidth=0.5,arrowsize=2 2,arrowlength=1.5]{->}(126,57)(131,55)
\psline[linewidth=0.5,arrowsize=2 2,arrowlength=1.5]{->}(126,70)(135,67)
\psline[linewidth=0.5,arrowsize=2 2,arrowlength=1.5]{->}(130,85)(142,78)
\rput(91,118){\System}
\rput(88,19){\Attack}
\psline[linewidth=0.5,linecolor=white,arrowsize=2 2,arrowlength=1.5]{->}(95,83)(97,92)
\psline[linewidth=0.5,linecolor=white,arrowsize=2 2,arrowlength=1.5]{->}(29,49)(41,54)
\psline[linewidth=0.5,linecolor=white,arrowsize=2 2,arrowlength=1.5]{->}(42,35)(49,40)
\psline[linewidth=0.5,linecolor=white,arrowsize=2 2,arrowlength=1.5]{->}(55,23)(59,32)
\psline[linewidth=0.5,linecolor=white,arrowsize=2 2,arrowlength=1.5]{->}(85,38)(77,39)
\psline[linewidth=0.5,linecolor=white,arrowsize=2 2,arrowlength=1.5]{->}(73,76)(67,81)
\psline[linewidth=0.5,linecolor=white,arrowsize=2 2,arrowlength=1.5]{->}(82,61)(73,61)
\psline[linewidth=0.5,linecolor=white,arrowsize=2 2,arrowlength=1.5]{->}(102,60)(109,60)
\psline[linewidth=0.5,linecolor=white,arrowsize=2 2,arrowlength=1.5]{->}(105,75)(112,80)
\psline[linewidth=0.5,linecolor=white,arrowsize=2 2,arrowlength=1.5]{->}(96,38)(103,38)
\psline[linewidth=0.5,linecolor=white,arrowsize=2 2,arrowlength=1.5]{->}(148,55)(137,59)
\psline[linewidth=0.5,linecolor=white,arrowsize=2 2,arrowlength=1.5]{->}(138,41)(132,45)
\psline[linewidth=0.5,arrowsize=2 2,arrowlength=1.5]{->}(109,33)(112,39)
\psline[linewidth=0.5,arrowsize=2 2,arrowlength=1.5]{->}(70,33)(67,38)
\end{pspicture}

%% file: PICS/4-adapt.tex
\ifx\JPicScale\undefined\def\JPicScale{1}\fi
\psset{unit=\JPicScale mm}
\psset{linewidth=0.3,dotsep=1,hatchwidth=0.3,hatchsep=1.5,shadowsize=1,dimen=middle}
\psset{dotsize=0.7 2.5,dotscale=1 1,fillcolor=black}
\psset{arrowsize=1 2,arrowlength=1,arrowinset=0.25,tbarsize=0.7 5,bracketlength=0.15,rbracketlength=0.15}
\begin{pspicture}(0,0)(159,131)
\newrgbcolor{userFillColour}{0.4 0 0}
\rput{0}(90,61){\psellipse[linewidth=1.15,fillcolor=userFillColour,fillstyle=solid](0,0)(50,-50)}
\newrgbcolor{userFillColour}{0.8 1 1}
\pscustom[linewidth=1.15,fillcolor=userFillColour,fillstyle=solid]{\psbezier(72,107.5)(68.5,106.45)(65.05,104.5)(60.5,101)
\psbezier(55.95,97.5)(52.65,94.2)(49.5,90)
\psbezier(46.35,85.8)(44.4,81.9)(43,77)
\psbezier(41.6,72.1)(40.85,68.35)(40.5,64.5)
\psbezier(40.15,60.65)(40.3,56.75)(41,51.5)
\psbezier(41.7,46.25)(43.65,41.45)(47.5,35.5)
\psbezier(51.35,29.55)(54.05,26.1)(56.5,24)
\psbezier(58.95,21.9)(61.95,19.95)(66.5,17.5)
\psbezier(71.05,15.05)(76,12.8)(83,10)
\psbezier(90,7.2)(88.95,5.1)(79.5,3)
\psbezier(70.05,0.9)(61.35,2.25)(50.5,7.5)
\psbezier(39.65,12.75)(33.35,18.3)(29.5,26)
\psbezier(25.65,33.7)(23.55,39.55)(22.5,45.5)
\psbezier(21.45,51.45)(21,57.3)(21,65)
\psbezier(21,72.7)(22.05,79)(24.5,86)
\psbezier(26.95,93)(29.95,98.25)(34.5,103.5)
\psbezier(39.05,108.75)(43.4,112.95)(49,117.5)
\psbezier(54.6,122.05)(60.3,125.05)(68,127.5)
\psbezier(75.7,129.95)(82.75,131)(91.5,131)
\psbezier(100.25,131)(107.45,129.5)(115.5,126)
\psbezier(123.55,122.5)(129.55,118.75)(135.5,113.5)
\psbezier(141.45,108.25)(145.5,103.6)(149,98)
\psbezier(152.5,92.4)(154.75,87)(156.5,80)
\psbezier(158.25,73)(159,67)(159,60)
\psbezier(159,53)(158.4,47.3)(157,41)
\psbezier(155.6,34.7)(153.05,29.3)(148.5,23)
\psbezier(143.95,16.7)(138.55,12.35)(130.5,8.5)
\psbezier(122.45,4.65)(116.6,2.7)(111,2)
\psbezier(105.4,1.3)(100.9,1.75)(96,3.5)
\psbezier(91.1,5.25)(92.75,7.65)(101.5,11.5)
\psbezier(110.25,15.35)(116.25,18.65)(121.5,22.5)
\psbezier(126.75,26.35)(130.5,30.7)(134,37)
\psbezier(137.5,43.3)(139.15,49.15)(139.5,56.5)
\psbezier(139.85,63.85)(139.25,69.7)(137.5,76)
\psbezier(135.75,82.3)(133.65,86.8)(130.5,91)
\psbezier(127.35,95.2)(124.35,98.2)(120.5,101)
\psbezier(116.65,103.8)(113.35,105.6)(109.5,107)
\psbezier(105.65,108.4)(102.95,109.3)(100.5,110)
\psbezier(98.05,110.7)(95.65,111)(92.5,111)
\psbezier(89.35,111)(86.35,110.7)(82.5,110)
\psbezier(78.65,109.3)(75.5,108.55)(72,107.5)
\closepath}
\psline[linewidth=0.5,linecolor=white,arrowsize=2 2,arrowlength=1.5]{->}(63,85)(55,93)
\psline[linewidth=0.5,linecolor=white,arrowsize=2 2,arrowlength=1.5]{->}(117,37)(125,29)
\psline[linewidth=0.5,linecolor=white,arrowsize=2 2,arrowlength=1.5]{->}(66,35)(58,27)
\psline[linewidth=0.5,linecolor=white,arrowsize=2 2,arrowlength=1.5]{->}(91,96)(91,108)
\psline[linewidth=0.5,linecolor=white,arrowsize=2 2,arrowlength=1.5]{->}(116,86)(125,93)
\psline[linewidth=0.5,linecolor=white,arrowsize=2 2,arrowlength=1.5]{->}(81,22)(78,15)
\psline[linewidth=0.5,linecolor=white,arrowsize=2 2,arrowlength=1.5]{->}(127,61)(137,61)
\psline[linewidth=0.5,linecolor=white,arrowsize=2 2,arrowlength=1.5]{->}(54,60)(43,60)
\psline[linewidth=0.5,linecolor=white,arrowsize=2 2,arrowlength=1.5]{->}(100,22)(103,15)
\rput(91,122){\System}
\rput(90,59){\Attack}
\psline[linewidth=0.5,arrowsize=2 2,arrowlength=1.5]{->}(150,38)(140,42)
\psline[linewidth=0.5,arrowsize=2 2,arrowlength=1.5]{->}(29,85)(40,80)
\psline[linewidth=0.5,arrowsize=2 2,arrowlength=1.5]{->}(61,9)(65,15)
\psline[linewidth=0.5,arrowsize=2 2,arrowlength=1.5]{->}(151,82)(140,78)
\psline[linewidth=0.5,arrowsize=2 2,arrowlength=1.5]{->}(64,120)(70,109)
\psline[linewidth=0.5,arrowsize=2 2,arrowlength=1.5]{->}(32,33)(44,39)
\psline[linewidth=0.5,arrowsize=2 2,arrowlength=1.5]{->}(122,11)(118,17)
\psline[linewidth=0.5,arrowsize=2 2,arrowlength=1.5]{->}(118,119)(111,109)
\end{pspicture}

%% file: 3-Conclusion.tex
\bigskip

\section{Final comments}\label{Summary}

\paragraph{On games of security and obscurity}
The first idea of this paper is that security is a game of \emph{incomplete} information: by analyzing your enemy's behaviors and algorithms (subsumed under what game theorists call his \emph{type}), and by obscuring your own, you can improve the odds of winning this game. 

This claim contradicts Kerckhoffs' Principle that there is no security by obscurity, which implies that security should be viewed as a game of \emph{imperfect} information, by asserting that security is based on players' secret data (e.g. cards), and not on their obscure behaviors and algorithms. 

I described a toy model of a security game which illustrates that security is fundamentally based on gathering and analyzing information about the type of the opponent. This model thus suggests that security is \emph{not} a game of imperfect information, but a game of \emph{incomplete} information. If confirmed, this claim implies that security can be increased not only by analyzing attacker's type, but also by obscuring defender's type. 

\medskip
\paragraph{On logical complexity}
The second idea of this paper is the idea of \emph{one-way programming}, based on the concept of \emph{logical complexity} of programs. The suggestion is that algorithmic information theory may be a useful new area for security research. It provides a natural conceptual framework for studying algorithm evolution driven by the battles between the attackers and defenders. It also provides some concrete technical tools, which raise a possibility of systems vulnerable to computationally feasible, but logically unfeasible attacks; in other words, the attack algorithms that may be easy to run when you know them, but hard to construct if you don't know them. Security experts dismiss such ideas, mainly because experience and theory show that algorithms are never too hard to \emph{re}\/construct from their obfuscations. But constructing them from other algorithms can be genuinely hard. 

\medskip
\paragraph{On security of profiling}
Typing and profiling are frowned upon in security. Leaving aside the question whether gathering information about the attacker, and obscuring the system, might be useful for security or not, these practices remain questionable socially. The false positives arising from such methods cause a lot of trouble, and tend to just drive the attackers deeper into hiding.

On the other hand, typing and profiling are technically and conceptually unavoidable in gaming, and remain respectable research topics of game theory. Some games cannot be played without typing and profiling the opponents.  Poker and the bidding phase of bridge are all about trying to guess your opponents' secrets by analyzing their behaviors. Players do all they can to avoid being analyzed, and many prod their opponents to sample their behaviors. Some games cannot be won by mere uniform distributions, without analyzing opponents' biases.

Both game theory and immune system teach us that we cannot avoid profiling the enemy. But both the social experience and immune system teach us that we must set the thresholds high to avoid the false positives that the profiling methods are so prone to. Misidentifying the enemy leads to auto-immune disorders, which can be equally pernicious socially, as they are to our health.

%% file: 4-Appendix.tex
\medskip
\section{Gaming security formalism}\label{Formalism}
\medskip

Can the idea of applied security by obscurity be realized? To test it, let us first make it more precise in a mathematical model. I first present a very abstract model of strategic behavior, capturing  and distinguishing the various families of games studied in game theory, and some families not studied. The model is based on coalgebraic methods, along the lines of \cite{PavlovicD:CALCO09}. I will try to keep the technicalities at a minimum, and the reader is not expected to know what is a coalgebra.

\medskip\subsection{Arenas}
\be{defn}
A \emph{player} is a pair of sets $A = <M_A, S_A>$, where the elements of $M_A$ represent or \emph{moves} available  to $A$, and the elements of $S_A$ are the \emph{states} that $A$ may observe.
 
A \emph{simple response} $\Sigma:A\to B$ for a player $B$ to a player $A$ is a binary relation 
\bear
\Sigma & : &    M_A\times S^2_B  \times  M_B \to \{0,1\}
\eear 
When $\Sigma(a,\beta, \beta', b)=1$, we write $<a,\beta>\tto \Sigma <\beta',b >$, and say that the strategy $\Sigma$ at $B$'s state $\beta$ prescribes that $B$ should respond to $A$'s move $a$ by the move $b$ and update his state to $\beta'$. The set of $B$'s simple responses to $A$ is written $\SR(A,B)$.

A \emph{mixed  response} $\Phi:A\to B$ for the player $B$ to the player $A$ is a matrix 
\bear
\Phi & : &     M_A\times S^2_B \times M_B  \to [0,1]
\eear 
required to be finitely supported and stochastic in $M_A$,  i.e. for every $a\in M_A$ holds
\begin{itemize}
\item $\Phi_{a\beta \beta' b } = 0$ holds for all but finitely many $\beta, \beta'$ and $b$, 
\item $\sum_{\beta \beta' b} \Phi_{a\beta \beta' b} = 1$.
\end{itemize}
When $\Phi_{a \beta\beta' b} = p$ we write $<a, \beta>\ttto \Phi p <\beta',b>$, and say that  the strategy $\Phi$ at $B$'s state $\beta$ responds to $A$'s move $a$ with a probability $p$  by $B$'s move $b$ leading him into the state $\beta' $. The set of $B$'s mixed responses to $A$ is written $\MR(A,B)$.

An \emph{arena} is a specification of a set of players and a set of responses between them.
\end{defn}

\paragraph{Responses compose} Given simple responses $\Sigma:A\to B$ and $\Gamma: B\to C$, we can derive a response $(\Sigma; \Gamma) : A\to C$ for the player $C$ against $A$ by taking the player $B$ as a "man in the middle". The derived  response is constructed as follows:
\[ \prooftree
<a, \beta > \tto \Sigma  <\beta',b> \qquad \qquad <b, \gamma >\tto\Gamma  <\gamma',c > 
\justifies
<a,\gamma >\tto{(\Sigma;\Gamma)} < \gamma',c >
\endprooftree\]
Following the same idea, for the mixed responses $\Phi:A\to B$ and $\Psi: B\to C$ we have the composite $(\Phi;\Psi):A\to C$ with
the entries
\bear
(\Phi;\Psi)_{a\gamma\gamma' c} & = & \sum_{ \beta\beta'b} \Phi_{a\beta\beta'b} \cdot \Psi_{b \gamma\gamma'c}
\eear
It is easy to see that these composition operations are associative and unitary, both for the simple and for the mixed responses.

\medskip\subsection{Games}
Arenas turn out to provide a convenient framework for a unified presentation of games studied in game theory \cite{VonNeumann-Morgenstern,Osborne-Rubinstein}, mathematical games \cite{winning-ways}, game semantics \cite{AJM,HO}, and some constructions in-between these areas \cite{Dubins-Savage}. Here we shall use them to succinctly distinguish between the various kinds of game with respect to the information available to the players. 
As mentioned before, game theorists usually distinguish two kinds of players' information:
\begin{itemize}
\item data, or positions: e.g., a hand of cards, or a secret number; and
\item types, or preferences: e.g., player's payoff matrix, or a system that he uses are components of his type.
\end{itemize}
The games in which the players have private data or positions are the games of \emph{imperfect information}. The games where the players have private types or preferences, e.g. because they don't know each other's payoff matrices, are the games of \emph{incomplete information}. See \cite{Aumann-Heifetz:handbook-incomplete} for more about these ideas, \cite{Osborne-Rubinstein} for the technical details of the perfect-imperfect distinction, and \cite{Fudenberg-Tirole:book} for the technical details of the complete-incomplete distinction.

But let us see how arenas capture these distinction, and what does all that have to do with security.

\medskip
\subsubsection{Games of perfect and complete information}
In games of perfect and complete information, each player has all information about the other player's data and preferences, i.e. payoffs. To present the usual stateless games in normal form, we consider the players $A$ and $B$ whose state spaces are the sets payoff bimatrices, i.e.
\bear
S_A\  =\  S_B & = & (\RRr \times \RRr)^{M_A \times M_B}
\eear
In other words, a state $\sigma \in S_A=S_B$ is a pair of maps $\sigma = <\sigma^A, \sigma^B>$ where $\sigma^A$ is the $M_A\times M_B$-matrix of $A$'s payoffs: the entry $\sigma^A_{ab}$ is $A$'s payoff if $A$ plays $a$ and $B$ plays $b$. Ditto for $\sigma_B$. Each game in the standard bimatrix form corresponds to one element  of both state spaces $S_A = S_B$. It is nominally represented as a state, but this state does not change. The main point here is that both $A$ and $B$ know this element. This allows both of them to determine the best response strategies $\Sigma_A:B\to A$ for $A$ and $\Sigma_B:A\to B$ for $B$, in the form
\bear
<b, \sigma^A> \tto{\Sigma_A} <\sigma^A, a> & \iff & \forall x\in M_A.\ \sigma^A_{xb}\leq \sigma^A_{ab}\\
<a,\sigma^B> \tto{\Sigma_B} <\sigma^B, b> & \iff & \forall y\in M_B.\ \sigma^B_{ay}\leq \sigma^B_{ab}
\eear
and to compute the Nash equilibria as the fixed points of the composites $(\Sigma^B; \Sigma^A):A\to A$ and  $(\Sigma^A; \Sigma^B):B\to B$. This is further discussed in \cite{PavlovicD:CALCO09}. Although the payoff matrices in games studied in game theory usually do not change, so the corresponding responses fix all states, and each response actually presents a method to respond in a whole family of games, represented by the whole space of payoff matrices, it is interesting to consider, e.g. discounted payoffs in some iterated games, where the full force of the response formalism over the above state spaces is used.

\medskip\subsubsection{Games of imperfect information}
Games of imperfect information are usually viewed in extended form, i.e. with nontrivial state changes, because players' private data can then be presented as their private states. Each player now has a set if private positions, $P_A$  and $P_B$, which is not visible to the opponent. On the other hand, both player's types, presented as their payoff matrices, are still visible to both. So we have
\bear
S_A & = & P_A \times (\RRr \times \RRr)^{M_A \times M_B}\\
S_B & = & P_B \times (\RRr \times \RRr)^{M_A \times M_B}
\eear
E.g., in a game of cards, $A$'s hand will be an element of $P_A$, $B$'s hand will be an element of $P_B$. With each response, each player updates his position, whereas their payoff matrices usually do not change.

\medskip\subsubsection{Games of incomplete information}
Games of incomplete information are studied in \emph{epistemic} game theory \cite{HarsanyiJ:bayesian,Mertens-Zamir,Aumann-Heifetz:handbook-incomplete}, which is formalized through knowledge and belief logics. The reason is that each player here only \emph{knows} with certainty his own preferences, as expressed by his payoff matrix. The opponent's preferences and payoffs are kept in obscurity. In order to anticipate opponent's behaviors, each player must build some \emph{beliefs} about the other player's preferences. In the first instance, this is expressed as a probability distribution over the other player's possible payoff matrices. However, the other player also builds beliefs about his opponent's preferences, and his behavior is therefore not entirely determined by his own preferences, but also by his beliefs about his opponent's preferences. So each player also builds some beliefs about the other player's beliefs, which is expressed as a probability distribution over the probability distributions over the payoff matrices. And so to the infinity. Harsanyi formalized the notion of players \emph{type} as an element of such information space, which includes each player's payoffs, his beliefs about the other player's payoffs, his beliefs about the other player's beliefs, and so on \cite{HarsanyiJ:bayesian}. Harsanyi's form of games of incomplete information can be presented in the arena framework by taking
\bear
S_A & = & \RRr^{M_A\times M_B} + \Delta S_B\\
S_B & = &  \RRr^{M_A\times M_B} + \Delta S_A
\eear
where $+$ denotes the disjoint union of sets, and $\Delta X$ es the space of finitely supported probability distributions over $X$, which consists of the maps $p:X\to [0,1]$ such that
\[ |\{x\in X|p(x) \gt 0\}|\lt \infty \quad \mbox{ and } \quad \sum_{x\in X} p(x) = 1\]
Resolving the above inductive definitions of $S_A$ and $S_B$, we get
\bear
S_A\  =\  S_B & = & \prod_{i=0}^\infty \Delta^{i} \left(\RRr^{M_A\times M_B}\right)
\eear
Here the state $\sigma^A\in S_A$ is thus a sequence $$\sigma^A = <\sigma^A_0, \sigma^A_1, \sigma^A_2,\ldots >$$ where $\sigma^A_i \in \Delta^i  \RRr^{M_A\times M_B}$. The even components $\sigma^A_{2i}$ represent $A$'s payoff matrix, $A$'s belief about $B$'s belief about $A$'s payoff matrix, $A$'s belief about $B$'s belief about $A$'s belief about $B$'s belief about $A$'s payoff matrix, and  so on. The odd components $\sigma_A^{2i+1}$ represent $A$'s  belief about $B$'s payoff matrix, $A$'s belief about $B$'s belief about $A$'s belief about $B$'s payoff matrix, and  so on. The meanings of the components of the state $\sigma^B\in S_B$ are analogous.

\input{4A-secgames.tex}

%% file: 4A-secgames.tex
\medskip\subsection{Security games}
We model security processes as a special family of games. It will be a game of imperfect information, since the players of security games usually have some secret keys, which are presented as the elements of their private state sets $P_A$ and $P_D$. The player $A$ is now the attacker, and the player $D$ is the defender.

The goal of a security game is not expressed through payoffs, but through \emph{"security requirements"} $\Theta \subseteq P_D$. The intuition is that the defender $D$ is given a family of assets to protect, and the $\Theta$ are the desired states, where these assets are protected. The defender's goal is to keep the state of the game in $\Theta$, whereas the attacker's goal is to drive the game outside $\Theta$. The attacker may have additional preferences, expressed by a probability distribution over his own private states $P_A$. We shall ignore this aspect, since it plays no role in the argument here; but it can easily be captured in the arena formalism.

Since players' goals are not to maximize their revenues, their behaviors are not determined by payoff matrices, but by their response strategies, which we collect in the sets $\Algs A D$ and $\Algs D A$. In the simplest case, response strategies boil down to the response maps, and we take $\Algs A D  =  \SR(A,D)$, or $\Algs A D = \MR(A,D)$. In general, though, $A$'s and $D$'s behavior may not be purely extensional, and the elements of $\Algs A D$ may be actual algorithms.

While both players surely  keep their keys secret, and some part of the spaces $P_A$ and $P_D$ are private, they may not know each other's preferences, and may not be given each other's "response algorithms". If they do know them, then both defender's defenses and attaker's attacks are achieved without obscurity. However, modern security definitions usually require that the defender defends the system against a family of attacks without querying the attacker about his algorithms. So at least the \emph{theoretical attacks are in principle afforded the cloak of obscurity}.  Since the defender $D$ thus does not know the attacker $A$'s algorithms, we model security games as games of incomplete information, replacing the player's spaces of payoff matrices by the spaces $\Algs A D$ and $\Algs D A$ of their response strategies to one another.

Like above, $A$ thus only knows $P_A$ and $\Algs D A$ with certainty, and $D$ only knows $P_D$ and  $\Algs A D$ with certainty. Moreover, $A$ builds his beliefs about $D$'s data and type, as a probability distribution over $P_D\times \Algs A D$, and $D$ builds similar beliefs about $A$. Since they then also have to build beliefs about each other's beliefs, we have a mutually recurrent definition of the state spaces again:
\bear
S_A & = & \big(P_A \times \Algs D A\big) + \Delta S_D \\
S_D & = & \big(P_D \times \Algs A D\big) + \Delta S_A
\eear
Resolving the induction again, we now get
\bear
S_A & = & \prod_{i=0}^\infty \Delta^{2i} \big(P_A\times \Algs D A\big) \times \Delta^{2i+1} \big(P_D\times \Algs A D\big)\\
S_D & = & \prod_{i=0}^\infty \Delta^{2i} \big(P_D\times \Algs A D \big) \times \Delta^{2i+1} \big(P_A\times \Algs D A\big)
\eear
Defender's state $\beta \in S_D$ is thus a sequence $\beta = <\beta_0, \beta_1, \beta_2,\ldots >$, where
\begin{itemize}
\item $\beta_0 = <\beta^P_{0}, \beta_0^\Al>\in P_D\times \Algs A D $ consists of
\begin{itemize}
\item $D$'s secrets $\beta^P_0\in P_D$, and
\item $D$'s current response strategy $\beta^\Al_0 \in \Algs A D  $
\end{itemize}
\item $\beta_1\in \Delta\big( P_A\times \Algs D A \big)$ is $D$'s belief about $A$'s secrets and her response strategy;
\item $\beta_2 \in \Delta^{2} \big(P_D\times \Algs A D\big)$ is $D$'s belief about $A$'s belief about $D$'s secrets and his response strategy;
\item $\beta_3\in \Delta^3 \big( P_A\times \Algs D A \big)$ is $D$'s belief about $A$'s belief about $D$'s beliefs, etc.
\end{itemize}
Each response strategy $\Sigma :A\to D$ prescribes the way in which $D$ should update his state in response to $A$'s observed moves. E.g., if $\Algs D A$ is taken to consist of relations in the form $\Lambda : {M_D^+ \times M_A}\to \{0,1\}$, where $M_D^+$ is the set of nonempty strings in $M_D$, then $D$ can record the longer and longer histories of $A$'s responses to his moves.

%
%
%
%
%
%
%
%
%
%
%

\paragraph{Remark} The fact that, in  a security game, $A$'s state space $S_A$ contains $\Algs D A$ and $\Algs A D$ means that each player $A$ is prepared for playing against a particular player $D$; while $D$ is prepared for playing against $A$. This reflects the situation in which all security measures are introduced with particular attackers in mind, whereas the attacks are built to attack particular security measures. A technical consequence is that players' state spaces are defined by the inductive clauses, which often lead to complex impredicative structures. This should not be surprising, since even informal security considerations often give rise to complex belief hierarchies, and the formal constructions of epistemic game theory \cite{HarsanyiJ:bayesian,Mertens-Zamir,FriedenbergA:hierarchy} seem like a natural tool to apply.

\medskip\subsection{The game of attack vectors}
We specify a formal model of the game of attack vectors as a game of perfect but incomplete information. This means that the players know each other's positions, but need to learn about each other's type, i.e. plans and methods. The assumption that the players know each other's position could be removed, without changing the outcomes and the strategies, by refining the way the model of players' observations. But this seems inessential, and we omit it for simplicity.

Let $\OOO$ denote the unit disk in the real plane. If we assume that it is parametrized in polar coordinates, then $\OOO = \{0\}\cup (0,1]\times \RRr/2\pi \ZZz$, where $\RRr/2\pi\ZZz$ denotes the circle. Let $\RRR\subseteq \RRr^2$ be an open domain in the real plane. Players' positions can then be defined as continuous mappings of $\OOO$ into $\RRR$, i.e.
\bear
P_A\  = \ P_D & = & \RRR^\OOO
\eear
The rules of the game prescribe that the attacker's and the defender's positions $p_A, p_D\in \RRR^\OOO$ maintain the invariant
\bear
 p^o_A \cap p^o_D & = & \emptyset
\eear
where $p^o$ denotes the interior of the image of $p:\OOO\to \RRR$. The assumption that each player knows both his own and his opponent's position means that both state spaces $S_A$ and $S_D$ contain $P_A\times P_D$ as a component. The state spaces are thus
\bear
S_A & = & (P_A\times P_D \times \Algs D A) + \Delta S_D\\
S_D & = & (P_A\times P_D \times \Algs A D) + \Delta S_D
\eear

To start off the game, we assume that the defender $D$ is given some assets to secure, presented as an area $\Theta\subseteq \RRR$. The defender wins as long as his position $p_D\in P_D$ is such that $\Theta \subseteq p^o_D$. Otherwise the defender loses.  The attacker's goal is to acquire the assets, i.e. to maximize the area $\Theta \cap p^o_A$.

The players are given equal forces to distribute along the borders of their respective territories. Their moves are the choices of these distributions, i.e.
\bear
M_A \ =\ M_D & = & \Delta\left(\partial \OOO\right)
\eear
where $\partial \OOO$ is the unit circle, viewed as the boundary of $\OOO$, and $\Delta (\partial \OOO)$ denotes the distributions along $\partial \OOO$, i.e. the measurable functions $m:\partial \OOO\to [0,1]$ such that $\int_{\partial \OOO} m= 1$. 

How will $D$ update his space after a move? This can be specified as a simple response $\Sigma: A \to D$. Since the point of this game is to illustrate the need for learning about the opponent, let us leave out players' type information for the moment, and assume that the players only look at their positions, i.e. $S_A = S_D = P_A\times P_D$. To specify $\Sigma: A \to B$, we must thus determine the relation
\bear
<m_A, p_A, p_D> \tto {\Sigma} <p'_A, p'_D, m_D>
\eear
for any given  $m_A \in M_A$,  $p_A\in P_A$, and $p_D\in P_D$. We describe the updates $p'_A$ and $p'_D$ for an arbitrary $m_D$, and leave it to $D$ to determine which $m_D$s are the best responses for him. So given the previous positions and both player's moves, the new position $p'_A$ will map a point on the boundary of the circle, viewed as a unit vector $\vec x \in \OOO$ into the vector $\vec p\,'_A(\vec x)$ in $\RRR\subseteq \RRr^2$ as follows.
\begin{itemize}
\item If for all $\vec y\in \OOO$ and for all $s\in \left[0,m_A(\vec x)\right]$ and all $t\in \left[0,m_D(\vec y)\right]$ holds $(1+s)\vec p_A(\vec x) \neq (1+t)\vec p_D(\vec y)$ then set
\bear
\vec p\, '_A(\vec x) & = & \frac{1+m_A(\vec x)}{2} \cdot \vec  p_A(\vec x) 
\eear
\item Otherwise, let $\vec y \in \OOO$, $s\in [0,m_A(\vec x)]$ and $t\in [0,m_D(\vec y)]$ be the smallest numbers such that $(1+s)\vec p_A(\vec x) = (1+t)\vec p_D(\vec y)$. Writing $m'_A(\vec x) = m_A(\vec x)-s$ and $m'_D(\vec y) = m_D(\vec y) -t$, we set 
\bear
\vec p\, '_A(\vec x) & = & \frac{1+m'_A(\vec x)}{2} \cdot \vec  p_A(\vec x)\ +\ \frac{1+m'_D(\vec y)}{2} \cdot \vec  p_D(\vec y) 
\eear
\end{itemize}
This means that the player $A$ will push her boundary by $m_A(\vec x)$ in the direction $\vec p_A(\vec x)$ if she does not encounter $D$ at any point during that push. If somewhere during that push she does encounter $D$'s territory, then they will push against each other, i.e. their push vectors will compose. More precisely, $A$ will push in the direction $\vec p_A(\vec x)$ with the force $m'_A(\vec x) = m_A(\vec x) - s$, that remains to her after the initial free push by $s$; but moreover, her boundary will also be pushed in the direction $\vec p_D(\vec y)$ by the boundary of $D$'s territory, with the force $m'_D(\vec y) = m_D(\vec y) - t$, that remains to $D$ after his initial free push by $t$.  Since $D$'s update is defined analogously, a common boundary point will arise, i.e. players' borders will remain adjacent. When the move next, there will be no free initial pushes, i.e. $s$ and $t$ will be 0, and the update vectors will compose in full force.

How do the players compute the best moves? Attacker's goal is, of course, to form a common boundary and to push towards $\Theta$, preferably from the direction where the defender does not defend. The defender's goal is to push back. As ex\-plained explained in the text, the game is thus resolved on defender's capability to predict attacker's moves. Since the territories do not intersect, but $A$'s moves become observable for $D$ along the part of the boundary of $A$'s territory that lies within the convex hull of $D$'s territory, $D$'s moves must be selected to maximize the length of the curve 
\[ \partial p_A\ \cap \ {\conv}(p_D)\]
This strategic goal leads to the evolution described informally in Sec.~\ref{Game}.